\documentclass{aastex}
\def\lsim{\raise0.3ex\hbox{$<$}\kern-0.75em{\lower0.65ex\hbox{$\sim$}}}
\def\gsim{\raise0.3ex\hbox{$>$}\kern-0.75em{\lower0.65ex\hbox{$\sim$}}}

\begin{document}

\title{Clustered Star Formation in Magnetic Clouds: \\
 Properties of Dense Cores Formed in Outflow-Driven Turbulence} 
\author{Fumitaka Nakamura\altaffilmark{1,2}, 
Zhi-Yun Li\altaffilmark{3}}
\altaffiltext{1}{National Astronomical Observatory, Mitaka, Tokyo 181-8588, 
Japan; fumitaka.nakamura@nao.ac.jp}
\altaffiltext{2}{Institute of Space and Astronautical Science, 
Japan Aerospace Exploration Agency, 3-1-1 Yoshinodai, Sagamihara, 
Kanagawa 229-8510, Japan}
\altaffiltext{3}{Department of Astronomy, University of Virginia,
P. O. Box 400325, Charlottesville, VA 22904; zl4h@virginia.edu}

%%%% limit of number of words is 250.
\begin{abstract}
We investigate the physical properties of dense cores formed in 
turbulent, magnetized, parsec-scale clumps of molecular clouds, 
using three-dimensional numerical simulations that include  
protostellar outflow feedback. The dense cores are identified 
in the simulated density data cube through a clumpfind algorithm. 
We find that the core velocity dispersion does not show any 
clear dependence on the core size, in contrast to Larson's 
linewidth-size relation, but consistent with recent 
observations. In the absence of a magnetic field, the majority 
of the cores have supersonic velocity dispersions. 
A moderately-strong magnetic field reduces the dispersion to  
a subsonic or at most transonic value typically. Most of the cores 
are out of virial equilibrium, with the external pressure 
dominating the self-gravity. The implication is that the core 
evolution is largely controlled by the outflow-driven turbulence. 
Even an initially-weak magnetic field can retard star formation 
significantly, because the field is amplified by the outflow-driven 
turbulence to an equipartition strength, with the distorted field 
component dominating the uniform one. In contrast, 
for a moderately-strong field, the uniform component remains dominant. 
Such a difference in the magnetic structure is evident 
in our simulated polarization maps of dust thermal emission; it 
provides a handle on the field strength. Recent polarization 
measurements show that the field lines in cluster-forming 
clumps are spatially well-ordered. It is indicative 
of a moderately-strong, dynamically important, field 
which, in combination with outflow feedback, can keep the rate 
of star formation in 
embedded clusters at the observationally-inferred, 
relatively-slow rate of several percent per free-fall time. 
\end{abstract}

\keywords{ISM: clouds --- ISM: magnetic fields ---
MHD --- polarization --- stars: formation --- turbulence}

\section{Introduction}
\label{sec:intro}

Millimeter and submillimeter observations of dense cores in nearby parsec-scale 
cluster-forming clumps have shown that the core mass function 
(CMF) resembles the stellar initial mass function (IMF).
For example, \citet{motte98} performed 1.2 mm continuum observations
toward L1688 and identified 57 starless dense cores whose mass function
($dN/dM \sim M^{-2.5}$) is consistent with the Salpeter power-law IMF 
at the high mass end ($dN/dM \sim M^{-2.35}$).
The CMF also has a break at about 0.3 M$_\odot$, below which it
flattens to $dN/dM \sim M^{-1.5}$.
Other authors have found similar CMFs in the region 
\citep[e.g.,][]{johnstone00,stanke06,maruta10}.
The shape of the CMF is broadly consistent with 
the IMF of Class II YSOs in the same region
\citep{bontemps01}. Another example is the Serpens
Cloud Core, where \citet{testi98} identified, through 3 mm continuum 
interferometric observations, 32 dense cores whose mass
function can be fitted by a power-law of $dN/dM \sim M^{-2.1}$,
again consistent with the Salpeter IMF. 
These and other observations suggest that the identified dense cores may be
the direct progenitors of individual stars and the bulk of the stellar 
IMF may be at least partly determined by cloud fragmentation
in the parsec-scale dense clumps.
Thus, understanding the formation process of dense cores is a key step
towards a full understanding of how stars form.

Both supersonic turbulence and magnetic fields are expected to play a
role in clustered star formation. The relative importance of the two 
is still under debated, however. One school of thought is that 
the core (and thus star) formation in a cluster-forming clump is
mainly regulated by supersonic turbulence. In this scenario, star 
formation is completed rapidly, before the initial supersonic 
turbulence has decayed significantly 
\citep[e.g.,][]{elmegreen00,hartmann01,klessen00}.  
Magnetic fields are less important in this picture, although 
\citet{padoan02} argued that a weak magnetic field is still needed 
in order to produce a core mass spectrum that resembles the stellar 
IMF. The weak magnetic field is expected to be strongly tangled by the
supersonic turbulence. Ordered field structures, if present at all,
are typically attributed to large-scale compression and are expected 
to be parallel to the dense filaments that are also produced by the 
same compression \citep[e.g.,][]{pelkonen07}.  

The second school of thought envisions a magnetic field that is more 
dynamically important; it prevents stars from forming too rapidly. 
Because supersonic turbulence dissipates rapidly, it must be
somehow replenished. 
In a cluster-forming clump, protostellar outflow feedback can play 
an important role in turbulence regeneration
\citep[e.g.,][]{maury09,arce10,nakamura11a,nakamura11b}.
\citet{li06} demonstrated that protostellar outflows 
can resupply the supersonic turbulence, keeping the clumps 
near a quasi-virial equilibrium state  
for a relatively long time \citep[see also][]{matzner07,nakamura07,carroll09}.
Because the matter moves preferentially along the dynamically
important magnetic field, the field is expected to be more or less  
perpendicular to the dense filaments that are created by either  
turbulence compression or self-gravity. A goal of this paper is 
to quantify the effects of the magnetic field on the clustered star 
formation in general, and the properties of dense cores in 
particular, through 3D MHD simulations that include outflow 
feedback for turbulence replenishment.

The rest of the paper is organized as follows. In Section
\ref{sec:model}, we describe the numerical method and simulation 
setup. In Section \ref{sec:results}, we apply a version of the 
so-called ``clumpfind'' algorithm to identify dense cores 
in a parent dense clump and derive their properties. We find 
that the internal motions of the cores are
sensitive to the magnetic field strength of the parent clump 
and that the external surface pressures play a dominant role 
in core formation in outflow-driven turbulence. 
In Section \ref{sec:discussion}, we compare the properties of 
the simulated cores with those observed in a couple of nearby 
cluster forming regions. 
Our main conclusions are summarized in Section \ref{sec:summary}.

\section{Model Formulation}
\label{sec:model}

The initial and boundary conditions of the simulations are 
essentially the same as those of \citet{nakamura07}. 
We consider a centrally-condensed molecular gas clump
inside a cubic simulation box of $L=1.5$~pc on each side,  
with an initial density profile of 
$\rho (r)=\rho_0 [1+(r/r_c)^2]^{-1}$ for $r\le L/2$ 
(where $r_c=L/6$ is the radius of the central plateau 
region) and  $\rho (r)=0.1 \rho_0$  for $r > L/2$. 
Here, the central density $\rho_0$ is given by 
$\rho_0=4.68\times 10^{-24} n_{\rm H_2,0}$ g cm$^{-3}$,
with $n_{\rm H_2,0}$ being the central number density of molecular
hydrogen, assuming 1 He for every 10 H atoms.
Periodic boundary condition is applied to each side of the box.
We adopt a central H$_2$ density of 
$n_{H_2,0} = 2.69\times 10^{4} (T/20 \ {\rm K})(1.5 \ {\rm pc}/L)^2$ cm$^{-3}$,
corresponding to a central free-fall time $t_{\rm ff,c} = 0.19$ Myr
and a central Jeans length of 
$L_J=(\pi c_s^2/G\rho_0)^{1/2} \simeq 0.17 (T/20 {\rm K})^{1/2}
(n_{\rm H_2,0}/2.69\times 10^4 {\rm cm}^{-3})^{-1/2}$ pc.
It yields a total clump mass of $M_{\rm tot} = 939 M_\odot$.
The average clump density is 
$\bar{n_{\rm H_2}}=4.0 \times 10^3$ 
cm$^{-3}$, corresponding to a global free-fall time 
$t_{\rm ff,cl}=0.49$ Myr. An isothermal equation of state 
is assumed, with a sound speed of $c_s = 0.266$ km s$^{-1}$
for gas temperature $T=$~20 K.

Our choice of the clump mass, density and gas temperature is 
motivated by observations of the nearest pc-scale cluster 
forming clump, the $\rho$ Ophiuchi main cloud.  
From $^{13}$CO ($J=1-0$) observations, \citet{loren89} 
estimated a total gas mass of 865~$M_\odot$ for the whole L1688 
region, adjusted for the parallax distance of 125~pc
\citep[e.g.,][]{lombardi08,loinard08}. 
The gas temperature in the region has some spatial variation, but 
the average appears close to 20~K. Our clump parameters are 
also consistent with those of other nearby cluster-forming clumps
\citep{ridge03} and the infrared dark clouds which are thought 
to be future sites of cluster formation in GMCs \citep[e.g.,][]{butler09}.

At the beginning of the simulation, we impose on the clump a 
uniform magnetic field along the $x$-axis. The field strength 
is specified by the plasma $\beta$, the ratio 
of thermal to magnetic pressures at the clump center, through
$B_0 = 47 \beta ^{-1/2} (T/20 {\rm K})^{1/2}(n_{\rm H_2,0} /2.69\times 
10^4 {\rm cm^{-3}} )^{1/2}$ $\mu$G.
In units of the critical value $1/ (2\pi G^{1/2})$
\citep{nakano78}, 
the mass-to-flux ratio in the central flux tube
is given by $\lambda_0 \simeq 8.3 \beta^{1/2}$.
The mass-to-flux ratio for the clump as a whole 
is $\bar{\lambda}\simeq 3.0\beta ^{1/2}$.
Although systematic measurements of magnetic field strength
in cluster-forming clumps are not available, 
relatively strong magnetic fields are inferred in some cases:
e.g., 850 $\mu$G for OMC 1 at $n\sim 10^{5-6}$ cm$^{-3}$ 
\citep[see][]{crutcher99, houde04,nakamura07} and
160 $\mu$G for Serpens at $n_{\rm H_2,0}\sim 10^{4-5}$ cm$^{-3}$ 
\citep{sugitani10}.
In the present paper, we adopt as a representative value $\beta = 0.2$ 
(or 100 $\mu$G at $n_{\rm H_2}\sim 3\times 10^4$ cm$^{-3}$), corresponding to 
$\bar{\lambda}\simeq 1.4$ or a dimensionless flux-to-mass ratio 
$\bar{\Gamma}\simeq 0.74$,
so that the clump as a whole, as well as the denser central 
region, is magnetically supercritical.

Following the standard practice \citep[e.g.,][]{maclow98,ostriker01},
we stir the initial clump at the beginning of the simulation with 
a turbulent velocity field of power spectrum $v_k \propto k^{-4}$ 
and rms Mach number ${\cal M}=5$. Our initial clump has a virial 
parameter (the
ratio of the kinetic energy to the gravitational energy) of 
$\alpha_{\rm vir} \simeq 0.5 $. The effective 
$\alpha_{\rm vir}$ should be close to unity for $\bar{\Gamma}\simeq
0.74$ since the gravitational energy is effectively reduced by the 
magnetic field by a factor of $1-\bar{\Gamma}^2$.
The turbulence is allowed to decay freely, except for feedback from
protostellar outflows.

The evolution of the turbulent, magnetized molecular clump
is followed using a three-dimensional MHD code based on an
upwind TVD scheme. 
The MHD code is essentially the same as those used
in \citet{li06} and \citet{nakamura07}. The ideal MHD equations
are solved using the code having second-order accuracy 
in both space and time.
The equation of state is assumed to be isothermal. 
To ensure the divergence-free magnetic field, the divergence 
cleaning method is adopted. The Poisson equation for gravitational
potential is solved using the fast Fourier transform.
Our simulation has a resolution of $256^3$. 
Although our focus is on core properties, we do include a 
crude treatment of the formed stars and outflows, following  
\citet{nakamura07}; we refer the reader to that paper
for details. Briefly, when the density in a cell crosses 
the threshold $\rho_{\rm th}=400 \rho_0$, corresponding to $10^7$ 
cm$^{-3}$, we create a Lagrangian particle at the cell 
center. We extract mass from a small region surrounding 
the cell, and put it on the particle, which moves with  
mass-weighted mean velocity of the extracted mass
(see Nakamura \& Li 2007 for more detail). After 
creation, the particle is allowed to accrete from the 
surrounding gas according to the same prescription 
as \citet{wang10}. 
To mimic the effect of protostellar 
outflows, the particle injects into the ambient gas a 
momentum that is proportional to the particle mass increment
$\Delta M_*$.  
The outflow momentum is scaled with the dimensionless 
outflow parameter $f$ as $P=f (\Delta M_{*}/M_\odot)(V_{\rm w}
/100{\rm km \ s^{-1}})$, where
$V_{\rm w}$ is the wind velocity 
\citep[see the Appendix of ][]{nakamura07}.
The fiducial value of $f$ is set to 0.4, consistent with 
\citet{matzner00}. Each outflow has a bipolar and spherical 
component, with a momentum ratio of 3:1. 
The model parameters are summarized in Table 
\ref{tab:model}.

\section{Numerical Results}
\label{sec:results}

We first concentrate on the dense cores identified in three 
representative models: model N1 (no magnetic field, $\beta=\infty$),  
W1 (weak magnetic field, $\beta=2$), and S1 
(moderately-strong magnetic field, $\beta = 0.2$). Their 
dimensionless outflow parameter is set to the fiducial value 
$f=0.4$. 

\subsection{Core Identification}

Following \citet{nakamura08}, we identify dense cores from the density data cube
using a variant of the CLUMPFIND algorithm of \citet{williams94}.
We adopt a threshold density $\rho_{\rm th,min}=4 \rho_0$ ($\sim 10^5$ 
cm$^{-3}$), and a maximum density $\rho_{\rm th, max}=400~\rho_{0}$
($\sim 10^7$ cm$^{-3}$, above which a Lagrangian particle is created). 
Thus, the cores identified from our simulation data correspond 
roughly to the dense cores observed in the dust continuum emission and
high density molecular tracers such as N$_2$H$^+$ ($J=1-0$) 
and H$^{13}$CO$^+$ ($J=1-0$). The density 
distribution between $\rho_{\rm th,min}$ and $\rho_{\rm th, max}$ is 
divided into 10 bins equally spaced logarithmically. 
We include only those spatially resolved cores containing more 
than 50 cells in the analysis, to ensure that their properties 
are determined with reasonable accuracy. 
We also choose the cores that do not include particles.
In this sense, our identified cores correspond to starless cores.
The minimum core mass and 
radius identified through this procedure are $M_{\rm c, min}\equiv 
\rho_{\rm th,min} \times 50 \Delta x \Delta y \Delta z \sim 0.05 
M_\odot$ and $\sim 0.01$~pc, respectively.
We note that our core identification procedure is not exactly the same
as those used in observational studies, which are typically based on 
position-position-velocity data cubes of molecular line emission or 
column density maps from dust continuum emission.

\subsection{Spatial Distribution of Dense Gas}

\begin{figure}
\plotone{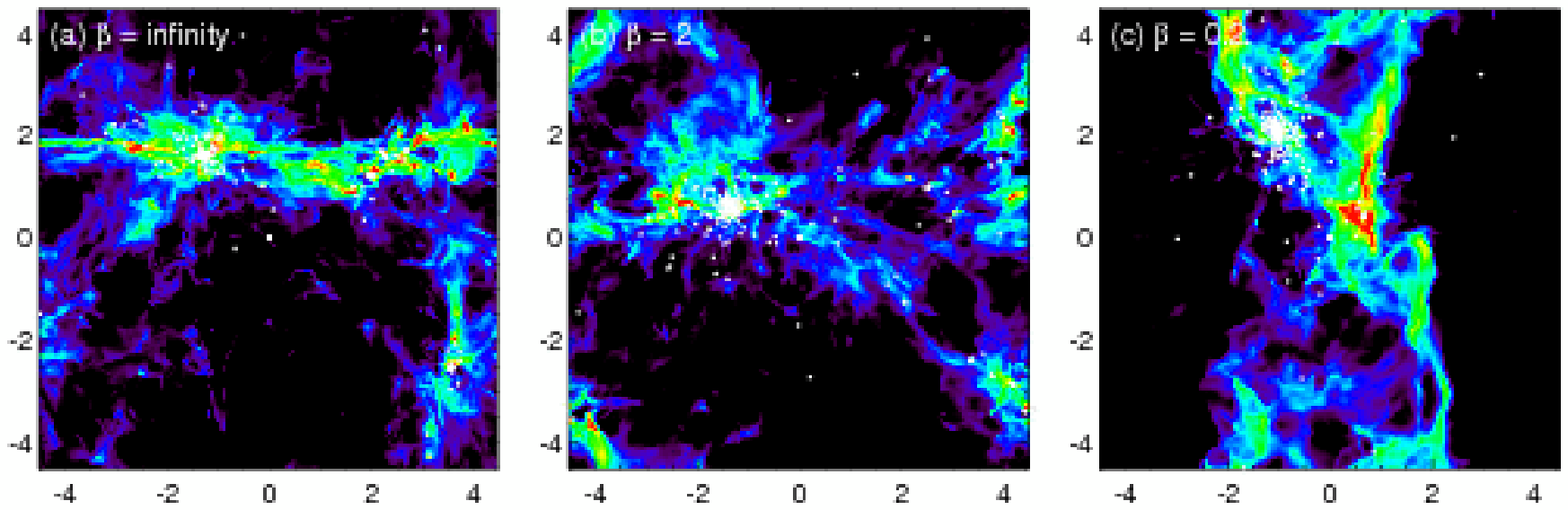}
\plotone{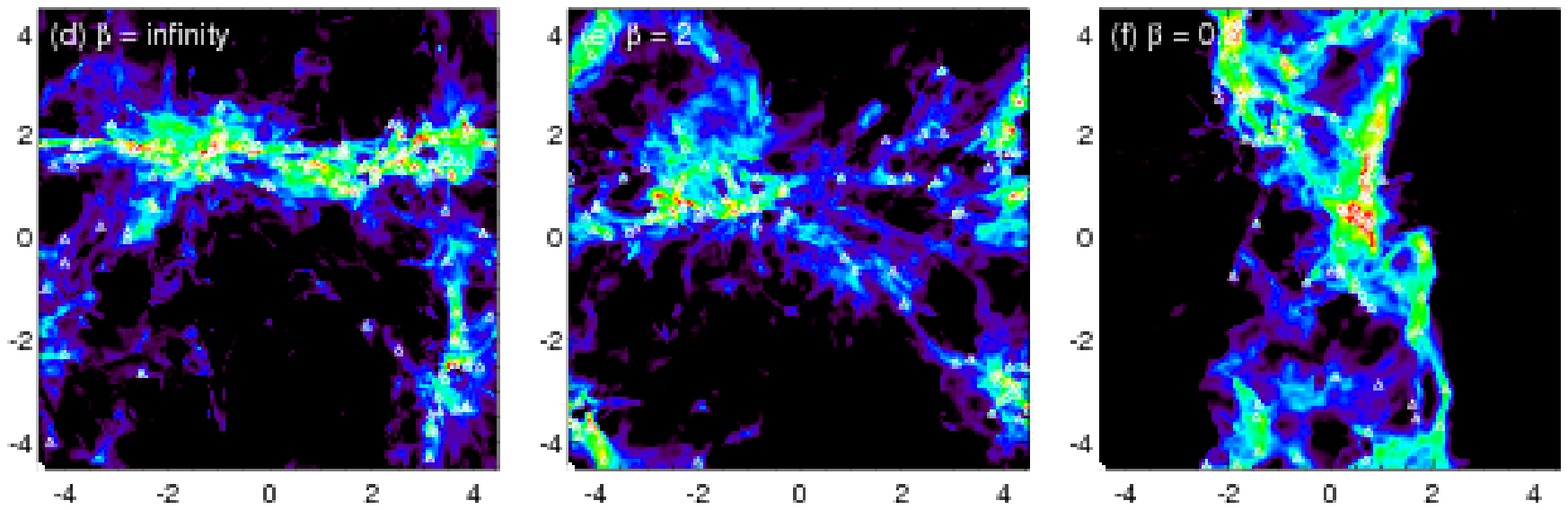}
\caption{Column density snapshots on the $x$-$z$ plane 
for (a) model N1 ($t=4.6 t_{\rm ff,c} \sim 0.9$ Myr), 
(b) W1 ($t=6.5 t_{\rm ff,c} \sim 1.2$ Myr), and 
(c) S1 ($t=7.4 t_{\rm ff,c} \sim 1.4$ Myr)
at the stage when 16 \% of the total mass has been
converted into stars.
The positions of the formed stars are overlaid with 
the dots.
The units of length are the central Jeans length
$L_J=(\pi c_s^2/G\rho_0)^{1/2} \simeq 
0.17 (T/20 {\rm K})^{1/2}(n_{\rm H_2,0}/2.69\times
 10^4 {\rm cm}^{-3})^{-1/2}$ pc.
The initial magnetic field direction is parallel to the $x$-axis.
In panels (d) through (f), the positions of the identified cores 
are overlaid with the triangles on the same column density images
as those shown in panels (a) through (c). 
}  
\label{fig:snapshot}
\end{figure}

Before discussing the physical properties of the identified cores, 
we show how the magnetic field affects the global density
distribution. Figure \ref{fig:snapshot} compares 
the column density distributions 
along the $y$-axis for the three models with different initial magnetic
field strengths, at a stage when 
the star formation efficiency has reached $16\%$. 
Since the star formation is retarded by the magnetic field, 
the evolution time tends to be longer at a given SFE for a stronger
magnetic field.
For comparison, the positions of the formed stars and the identified 
dense cores projected on the $x-z$ plane
are indicated by the dots and triangles, respectively, 
in Fig. \ref{fig:snapshot}.
The initial magnetic field direction is parallel 
to the $x$-axis (the abscissa).

Figure \ref{fig:snapshot} indicates that the global density distribution
depends on the initial magnetic field strength.
For the non-magnetic model (model N1), 
the global density distribution 
shows a large-scale filamentary structure that contains
many small fragments (Figs. \ref{fig:snapshot}a and \ref{fig:snapshot}d).
The formation of the large-scale filament is probably because the 
turbulence energy is initially highest on the largest scale and the
large-scale compression may be further amplified by self-gravity to 
create the filamentary structure.

For the weakly-magnetized case (model W1), the global density distribution 
does not clearly show a large-scale filamentary structure 
by the stage shown in Figs. \ref{fig:snapshot}b and \ref{fig:snapshot}e.
The dense part appears to contain many smaller filaments or elongated 
fragments that are distributed almost independently 
of the initial magnetic field direction.
This is different from the non-magnetic case. The difference may be
because the
local turbulent motions amplify random magnetic field components
(see Section \ref{subsec:amplification}),
which slow down the formation of the large-scale filamentary structure.

In contrast, in the presence of a moderately-strong magnetic field
(model S1),
the cloud material condenses preferentially along the magnetic field 
lines into a large-scale filamentary structure that is nearly  
perpendicular to the initial magnetic field direction.
Several less-dense filaments are also associated with the
large-scale filament (hereafter the main filament).
The less-dense filaments appear to be more or less along
the initial magnetic field direction and to merge toward the dense parts 
of the main filament.  
The dense cores are distributed primarily along the main filament.
The column density distribution is reminiscent of the 
hub-filament structures of star forming regions pointed out
by \citet{myers09}. The observed hub-filament structures may  
indicate the presence of the moderately-strong, spatially-ordered 
magnetic field, instead of the weak magnetic field that are 
tangled by supersonic turbulence.
We will revisit this issue in Section \ref{subsec:polarization}, where
we compare dust polarization maps derived from the simulation 
data for models W1 and S1.

\subsection{Core Property}

As shown in \citet{li06} and \citet{nakamura07},
our model clump has reached a quasi-equilibrium state 
within one global free-fall time.
After that, the statistical properties of the cores 
remain largely-unaltered.
Therefore, to compare the statistical properties of the identified
cores, we choose the same stage as that of Fig. \ref{fig:snapshot},
i.e., the stage when 16 \% of the total mass has been
converted into stars.
In this subsection, we compare the core properties 
for the three representative models, models N1, W1, and S1.
The total numbers of identified cores are 160, 105, and 105 cores for models 
N1, W1, and S1, respectively.

\subsubsection{Radius, Mass, and Velocity Dispersion}

\begin{figure}
\epsscale{1.0}
\plotone{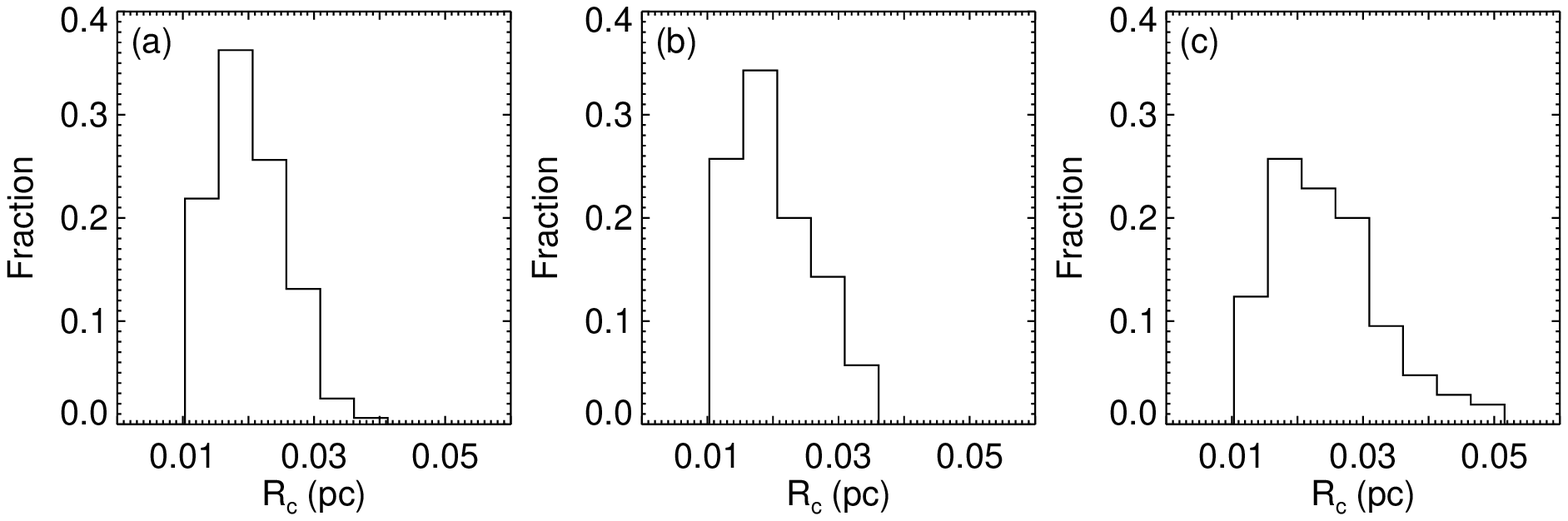}
\plotone{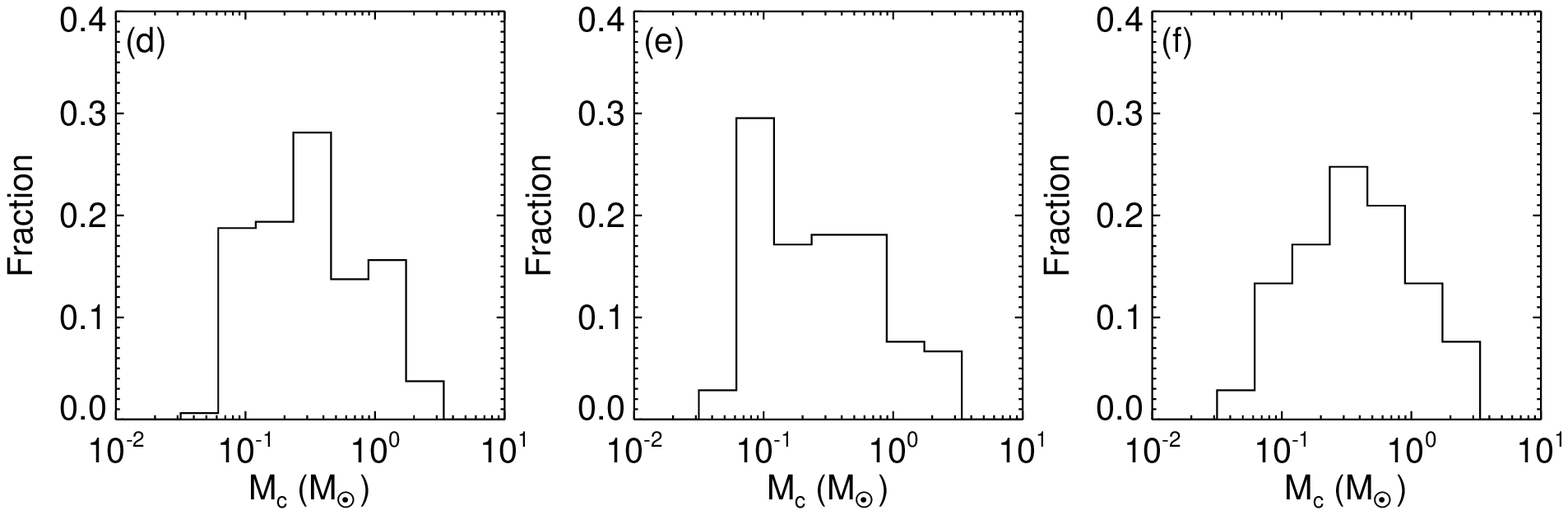}
\plotone{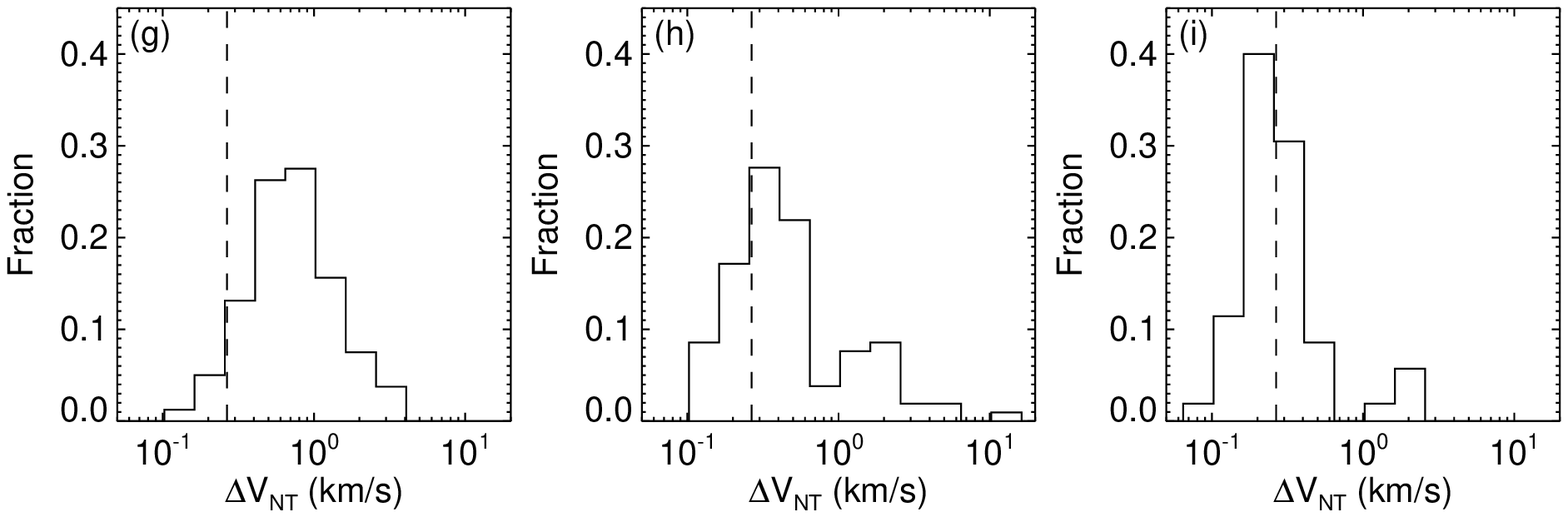}
\caption{Histograms of some physical quantities of the identified cores.
They are plotted at the same stages as those of Fig. \ref{fig:snapshot}.
Panels (a), (b), and (c) are the histograms of the core radius for 
models N1, W1, and S1, respectively.
Panels (d), (e), and (f) are the histograms of the core mass for 
models N1, W1, and S1, respectively.
Panels (g), (h), and (i) are the histograms of the 3D velocity dispersion
of the nonthermal component for models N1, W1, and S1, respectively.
The vertical dashed lines indicate the sound speed with 
gas temperature of 20 K.
}  
\label{fig:hist}
\end{figure}

We now focus on the physical properties of the cores identified in 
the three representative models.  
To facilitate comparison with observations, 
we present the core properties in dimensional units.
Figures \ref{fig:hist}a through \ref{fig:hist}i 
show the histograms of the core
radius, mass, and velocity dispersion of the nonthermal component.
The minimum, maximum, mean, and median values are summarized in
Table \ref{tab:physprop}.  
The core radius, $R_{\rm c}$, is defined as the 
radius of a sphere having the same volume as the core.
The distributions of core radius are broadly similar in all cases. 
They range from $\sim$0.01~pc to $\sim $0.05~pc, all peaking 
around 0.02~pc. The maximum and mean core radii are somewhat 
larger in model S1 (the moderately-strong field case) than in 
the other two. The core mass distribution shows a similar trend.
In all models, the core mass ranges from $\sim$ 0.07 $M_\odot$ 
to $\sim 4 M_\odot$. The mean value is somewhat larger for 
model S1 than for the other two.
The mean core mass and radius are larger in model S1 probably because 
the stronger magnetic field can support more mass against gravitational 
collapse.

The magnetic effect is even more prominent in the distribution of core 
velocity dispersion. 
For the non-magnetic model (model N1), the velocity dispersion ranges from 
$\sim$~0.1 km s$^{-1}$ to 4 km s$^{-1}$, 
peaking around 1 km s$^{-1}$ (corresponding to a Mach number of $\sim $ 4). 
Most of the cores have supersonic internal motions in this case, which 
appears inconsistent with the predominantly subsonic or at most 
transonic internal motions inferred observationally 
(e.g., Andre et al. 2007; Walsh et al. 2007; Maruta et al. 2010, 
see also Section \ref{subsec:observations}).
Even a relatively weak magnetic field appears to affect the 
internal motions inside the cores significantly.
For model W1, the median velocity dispersion is estimated at $\sim$0.5~km~s$^{-1}$, corresponding to a Mach number of about 2, although 
the mean is almost the same as that of model N1.
Most of the cores in this case still have supersonic velocity dispersions.

In contrast, for model S1, the majority of the cores have a velocity 
dispersion close to, or less than, $\sim$ 0.3 km s$^{-1}$, which is 
comparable to the sound speed ($c_s=0.266 $ km s$^{-1}$). The velocity 
dispersion ranges from $\sim$~0.05 km s$^{-1}$ to 2 km s$^{-1}$, peaking 
around the median of 0.3 km s$^{-1}$. In other words, the strong 
magnetic field tends to reduce the core internal motions significantly.
The reduction is probably because the magnetic field 
cushions the converging flows in the supersonic turbulence that are 
largely responsible for the core formation. The magnetic effect is 
also evident in the virial state of the cores, as we show below.

\subsubsection{Velocity Dispersion-Radius Relation}

The internal motions of the observed dense cores are often measured 
using the velocity dispersions derived from molecular line data. 
Here, we compare the velocity dispersions for the three representative 
models with different field strengths to study the effects of the 
magnetic field.

\begin{figure}
\plotone{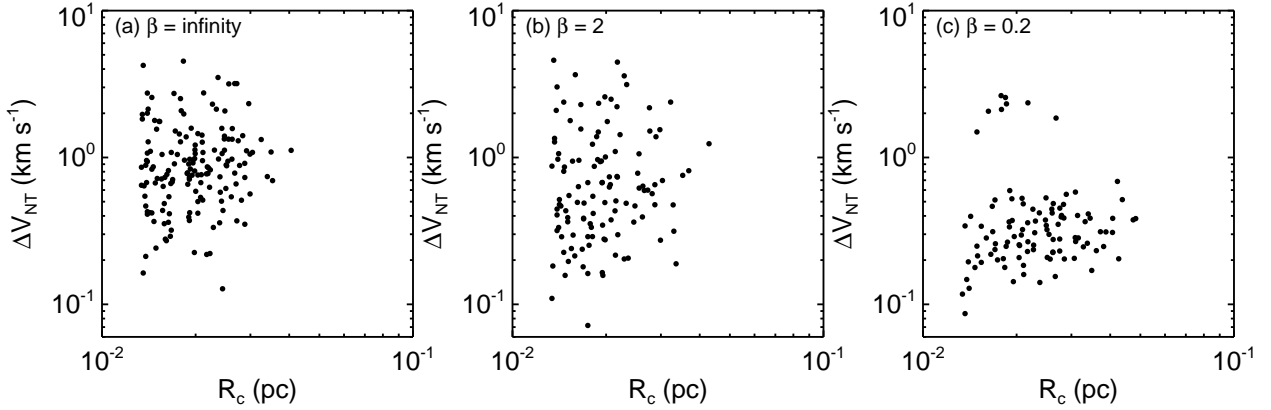}
\caption{Nonthermal 3D velocity dispersions as a function of core radii
for three models with different magnetic field strengths:
(a) model N1, (b) model W1, and (c) model S1. 
They are plotted at the same stages as those of Fig. \ref{fig:snapshot}.
The sound speed is $c_s=0.266$ km s$^{-1}$, corresponding to $T=20$ K. 
}  
\label{fig:velocity-radius}
\end{figure}

Figure \ref{fig:velocity-radius} plots the non-thermal components of
3D velocity dispersions of the identified cores against the core radii 
for (a) model N1, (b) model W1, and (c) model S1. 
The best-fit power-laws are given by 
$(\Delta V_{\rm NT}/ {\rm km \ s^{-1}})=(2.2 \pm 1.9)
(R_{\rm c}/{\rm pc})^{0.19\pm 0.22}$ (${\cal R} = 0.023$),
$(\Delta V_{\rm NT}/ {\rm km \ s^{-1}})=(3.8\pm 9.0)
(R_{\rm c}/{\rm pc})^{0.35\pm 0.61}$ (${\cal R} = 0.070$), and
$(\Delta V_{\rm NT}/ {\rm km \ s^{-1}})=(1.1 \pm 0.48)
(R_{\rm c}/{\rm pc})^{-0.34\pm 0.12}$ (${\cal R} = 0.29$)
for models N1, W1, and S1, respectively, where
${\cal R}$ denotes the correlation coefficient.
It shows that there 
is no clear correlation between the velocity dispersion and core 
radius. Our cores do not follow the power-law linewidth-radius 
relation derived by \citet{larson81}. The same conclusion holds for  
dense cores formed in even more strongly magnetized, magnetically 
subcritical, clouds \citep[see Figure 18 of][]{nakamura08}. The 
reason is probably that the dense cores created out of turbulent 
molecular gas are generally not in virial equilibrium (the equilibrium 
is implied by Larson's laws) and that their formation is strongly 
influenced by external turbulent flows (see below).

Quantitatively, the velocity dispersions of the cores depend 
on the initial magnetic field strength, as shown already in Figs. 
\ref{fig:hist}g through \ref{fig:hist}i.
In the absence of magnetic field (model N1), 
about 40\% of the cores have velocity dispersions larger than 1 km
s$^{-1}$; the majority of the cores have supersonic internal motions.
As mentioned above, even a relatively weak magnetic field can 
influence the internal motions significantly.  In model W1, although 
many cores still have large velocity dispersions, the median 
value is less than that in model N1 by a factor of $\sim 2$. 
In the stronger field case (model S1), the median and mean of the 
velocity dispersions are estimated at 0.31 and 0.46 km s$^{-1}$. 
About 40 \% of the cores have velocity dispersions smaller than 
the sound speed of 0.266 km s$^{-1}$; the majority of the cores 
have subsonic or at most transonic internal motions, which 
is more consistent with observations. We conclude that the magnetic 
field plays a significant role in reducing the internal motions of 
the cores.

\subsubsection{Virial Parameter}

\begin{figure}
%\epsscale{0.8}
\plotone{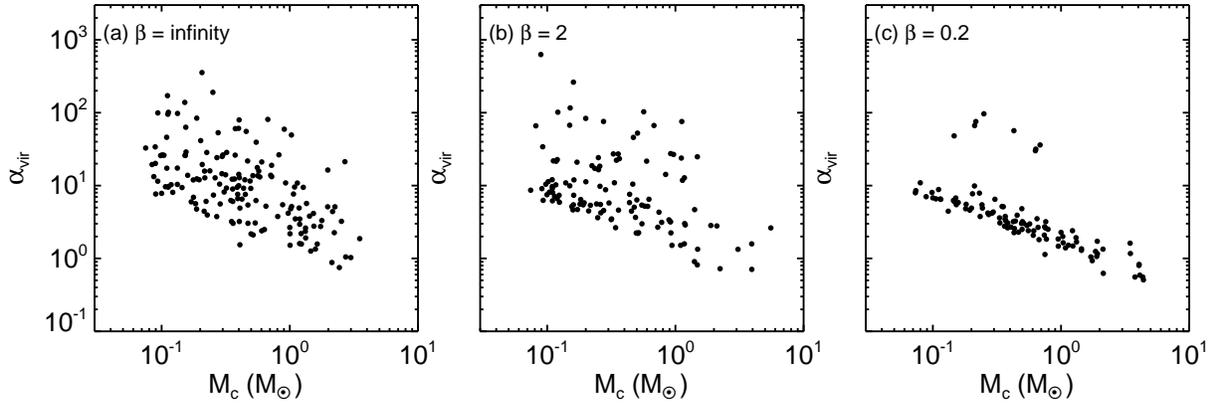}
\caption{Virial parameters as a function of core masses
for three models with different magnetic field strengths:
(a) model N1, (b) model W1, and (c) model S1. 
They are plotted at the same stages as those of 
Fig. \ref{fig:velocity-radius}.
}  
\label{fig:virial parameter}
\end{figure}

The virial parameter, the ratio of the virial mass to core mass, 
is often used as a measure of the gravitational boundedness
of dense molecular cloud cores, particularly in observational studies. 
In Fig. \ref{fig:virial parameter}, we plot the virial parameters
against the core mass for the three models.
Here, we calculate the virial parameter as
\begin{equation}
\alpha_{\rm vir}= \frac{5\Delta V_{\rm 1D}^2 R_{\rm c}}{GM_{\rm c}}  \  , 
\end{equation}
where $\Delta V_{\rm 1D}$ is the 1D FWHM velocity width 
including thermal contribution and 
the virial mass of each core is obtained by assuming 
that it is a uniform sphere.
The virial mass depends on the density distribution.
For a centrally condense sphere with $\rho \propto r^{-2}$,
the virial parameter is reduced by a factor of $5/3$.
Note that the effects of the nonspherical mass distribution 
appear to be small as long as the aspect ratios of the cores are not  
far from unity \citep{bertoldi92}.

Figure \ref{fig:virial parameter} indicates that the relationship 
between the virial parameter and core mass
depends on the initial magnetic field strength.
In the absence of a magnetic field (model N1), the virial 
parameters are typically very large, and have a large scatter. 
The scatter in the virial parameter-mass plot 
comes mainly from the large scatter in the velocity dispersion-radius relation 
(see Fig. \ref{fig:velocity-radius}).
The virial parameters often 
exceed $10 - 10^2$ even for relatively massive cores, 
indicating that, for most of the cores, 
the gravitational energy is much smaller than the internal kinetic
energy. 
The virial parameters tend to increase with decreasing core mass, and 
the lower bound of the virial parameters follows roughly  
a power-law relation of $\alpha_{\rm vir} \propto M_{\rm c}^{-2/3}$.

For the model with a weak magnetic field (model W1), the virial parameter-mass 
relation is qualitatively similar to that of the non-magnetized model
(model N1). 
Although the fraction of cores with large virial parameters is
somewhat small for the weak magnetic field case, the scatter in 
the viral parameter-mass relation remains large. 
About 30 \% of the cores still have virial parameters
larger than 10.

On the other hand, for the model with a moderately-strong magnetic
field (model S1),
the identified cores tend to follow a single power-law
relation of $\alpha_{\rm vir} \propto M_{\rm c}^{-2/3}$, 
with a rather small scatter.  
Furthermore, most of the cores (about 9 \% of the cores) 
have virial parameters smaller than 10. 
This trend is consistent with the initially-magnetically subcritical
case \citep[see Figure 17 of][]{nakamura08}.
It is also in good agreement with the scaling found by
\citet{bertoldi92} for nearby giant molecular clouds.
According to \citet{bertoldi92}, an object confined by the ambient pressure 
has a virial ratio that is proportional to $(M_{\rm c}/M_J)^{-2/3}$, 
where $M_J$ is the Jeans mass defined by Equation (2.13) of \citet{bertoldi92}.
The fact that our cores follow the same relation implies that the 
ambient pressure plays an important role in confining our cores as well. 
In the next subsection, we perform a detailed virial analysis 
of the cores to quantitatively clarify the role 
of the surface pressure term in core dynamics.

\subsubsection{Virial Analysis}

The virial theorem is useful for analyzing the core dynamical
properties. Here, we perform a detailed virial analysis
using the virial equation in Eulerian coordinates
\citep[e.g.,][]{mckee92,tilley07,dib07} 
\begin{equation}
S=\frac{1}{2}\ddot{I} + \frac{1}{2}\frac{d}{dt}\int _S \rho r^2 
\mbox{\boldmath $v$} \cdot \mbox{\boldmath $dS$}
= U+K+M+W+S+F \ ,
\label{eq:virial}
\end{equation}
where $I=\int_V \rho r^2 dV$, $U=3 \int_V P dV$, 
$K=\int_V \rho v^2 dV$, 
$W=-\int_V \rho \mbox{\boldmath $r$} \cdot \nabla \Psi dV$, 
$M=(1/8\pi)\int _V B^2 dV$, 
$S= - \int _S [P\mbox{\boldmath $r$}+\mbox{\boldmath $r$}\cdot
(\rho \mbox{\boldmath $r$}\mbox{\boldmath $r$})] \cdot 
\mbox{\boldmath $dS$}$, and 
$F=\int _S \mbox{\boldmath $r$}\cdot \mbox{\boldmath $T_M$} 
\cdot \mbox{\boldmath $dS$}$.
The quantities $I$, $U$, $K$, $W$, $S$, 
$M$, $F$, and $\mbox{\boldmath $T_M$}$ 
denote, respectively, the moment of inertia, 
internal thermal energy, internal kinetic energy, gravitational energy, 
the sum of the thermal surface pressure and dynamical surface pressure, 
internal magnetic energy, the magnetic surface pressure, and the Maxwell
stress-energy tensor. 
The internal thermal, kinetic, and magnetic energies are always
positive. Other terms can be either positive or negative.
In particular, the gravitational term can be positive in
crowded environments where the background gravitational 
field dominates the core dynamics \citep[see e.g.,][]{ballesteros09}.
In fact, as shown below, several cores have positive gravitational terms
in our simulations, although no cores have positive gravitational terms
for the initially magnetically subcritical case where
the turbulent motions are not so strong because of the effects of 
magnetic cushion 
\citep[see Fig. 16 of][]{nakamura08}.
The second term on the left-hand side of equation (\ref{eq:virial}) 
denotes the time derivative of the flux of moment of inertia 
through the core boundary. As is the standard practice 
\citep{tilley07, dib07,nakamura08}, we ignore the left-hand side of 
equation (\ref{eq:virial}) in our discussion and consider a core 
to be in virial equilibrium if the sum $U+K+W+S+M+F = 0$.

\begin{figure}
%\epsscale{0.8}
\plotone{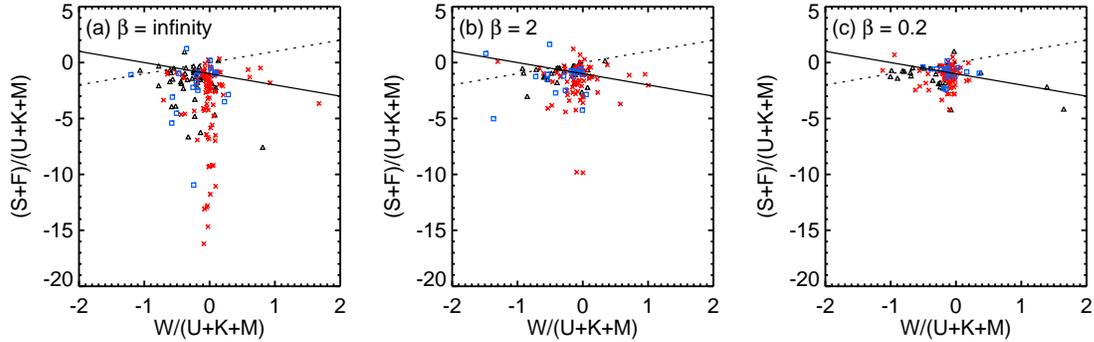}
\caption{
Relationship between the sum of the two surface terms, $S+F$, 
and the gravitational term, $W$, in the virial equation. 
They are normalized to the sum of the internal terms, $U+K+M$. 
The solid line indicates the virial equilibrium,
$U+K+W+S+M+F=0$. 
They are plotted at the same stages as those of 
Fig. \ref{fig:velocity-radius}.
The dashed line indicates the line below which the surface term is 
larger than the gravitational term. 
For the cores that lie below the solid line, the left-hand side of the
virial eq. \ref{eq:virial} is negative and thus expected to be bound.
All others that lie above the solid line are unbound and expected to disperse
away, if they do not gain more mass through accretion and/or merging 
with other cores, or reduce internal support through turbulence dissipation.
A red cross, black triangle, and blue square indicates a core with mass of 
$M_{\rm c} \le 0.75 \left<M_{\rm c}\right>$, 
$0.75 \left<M_{\rm c}\right> \le M_{\rm c} 
\le 1.25 \left<M_{\rm c}\right>$,
and $M_{\rm c} \ge 1.25 \left<M_{\rm c}\right>$, respectively.
The close-up of Fig. \ref{fig:virial analysis}c is presented in 
Fig. \ref{fig:outflow2}c.
}  
\label{fig:virial analysis}
\end{figure}

The equilibrium line is shown in Figure \ref{fig:virial analysis}, where
the sum of the surface terms ($S+F$) is plotted against the
gravitational term ($W$), both normalized to the sum of the internal
terms ($U+K+M$).  For the surface term, the external kinetic term 
is generally much larger than the magnetic term ($F$).

For all three models, the majority of the cores lie below the virial
equilibrium line, and thus they are expected to be unstable 
to contraction.  
In addition, the surface term appears more important than the
gravitational term for most of the cores.  
In the absence of the magnetic field, about 10 $-$ 20 \% of the cores
have the normalized surface term of $|S+F|/|U+K+M| \gtrsim 5$, 
whereas no cores have the normalized surface term of $|S+F|/|U+K+M|
\gtrsim 5$ in the presence of the moderately-strong magnetic field.
The magnetic field tends to reduce the contribution of the surface
terms to the core dynamical state, although the surface term 
is still dominant even in the presence of the strong magnetic field.

The core virial state is in contrast to the 
initially-magnetically subcritical case.
For the initially-magnetically subcritical case, 
for most of the cores, the surface term ($S+F$) and 
gravitational term ($W$) tend to be smaller than the sum of 
the internal terms, $U+K+M$ \citep{nakamura08}, and thus 
most of the cores are distributed around the virial equilibrium line.
In addition, more massive cores tend to be distributed below the 
virial equilibrium line, and 
the importance of the self-gravity tends to depend on the core mass.  
The gravitational term tends to be more important for 
more massive cores for the initially magnetically
subcritical case, implying that the gravitational collapse 
induced by ambipolar diffusion regulates the core evolution.  
In contrast, for the magnetically-supercritical
cases studied in this paper, most of the cores are distributed 
further from the virial equilibrium line.
The surface term ($S+F$) tends to be larger than the sum of 
the internal terms, $U+K+M$ for most of the cores.
The core virial state shows no clear dependence on the mass.  
The surface term, which is typically dominated by the ambient 
supersonic turbulence, is still important even for massive cores.

The above result is different from that of \citet{tilley07}. They 
found, in their ideal MHD simulations, a larger contribution from 
the surface term to the virial equation for a stronger magnetic 
field, which is oppose to the trend we found. 
This difference likely comes from the fact that 
they only followed the very early phase of cloud evolution, when  
the initial supersonic turbulence has decayed significantly but 
star formation has yet to set in. Most of our cores are formed at
much later times, after the decayed turbulence has been replenished by 
outflow feedback. Given this difference, it is perhaps not too 
surprising that the surface terms are much larger in our models 
than in theirs. This is particularly true for the weaker magnetic 
field case, because of a more rapid and violent star formation.

In our simulations, most of the identified cores
have the densities smaller than the critical density beyond 
which the Jeans condition is violated. For example, 
the Jeans condition is satisfied for 89 \% (model N1), 85 \%
(model W1), and 95 \% (model S1) of the cores presented in this subsection.
For the stronger initial magnetic field, the mean density 
of the cores tend to be smaller due to the magnetic support, and 
therefore the Jeans condition is satisfied for a larger number of cores.
We also performed a run with 512$^3$ for model S1, 
with the threshold density fixed, so that the Jeans condition is 
always satisfied in the entire computation box.
To compare directly with the results presented in this subsection,
we first resized the data obtained from the $512^3$ run to
the $256^3$ grid data, and then applied the clumpfind method to the
regridded data. 
We confirmed that the statistical properties of the identified cores
are essentially the same as those presented in this subsection.
For example, for all the cores identified from the $512^3$ data, 
the nonthermal 3D velocity dispersions of the identified cores stay 
below 1 km s$^{-1}$, and its mean is estimated to be 0.43 km s$^{-1}$,
comparable to that of the $256^3$ data, 0.46 km s$^{-1}$.

\subsection{Effects of Outflow Strength}

After exploring the effects of the magnetic field on the core
properties, we now turn to the outflow strength. 
In our model, the strength of the outflow feedback is specified 
by the dimensionless parameter, $f$.
This parameter is highly uncertain because it is generally difficult
to determine the physical properties of molecular outflows accurately
from observational data. In this paper, we adopt $f=0.4$ as a fiducial value,
following \citet{matzner00}. Recently, Nakamura et al. (2011)
discovered many molecular outflow lobes toward an 
extremely-young cluster-forming clump, Serpens South, 
on the basis of CO $(J=3-2)$ observations, and 
derived the physical properties of the identified molecular outflows.
They roughly estimated  $f\sim 0.1 - 0.3$ for these outflows,
assuming that the outflow gas is optically-thin.
If a substantial amount of the outflow gas is optically-thick,
the outflow parameter $f$ would be larger.
In the following, we compare the core properties of three models with
different outflow strengths: $f=0$, 0.1, and 0.4. The fiducial value
for plasma $\beta$ of 0.2 is adopted for these models.

\begin{figure}
%\epsscale{0.8}
\plotone{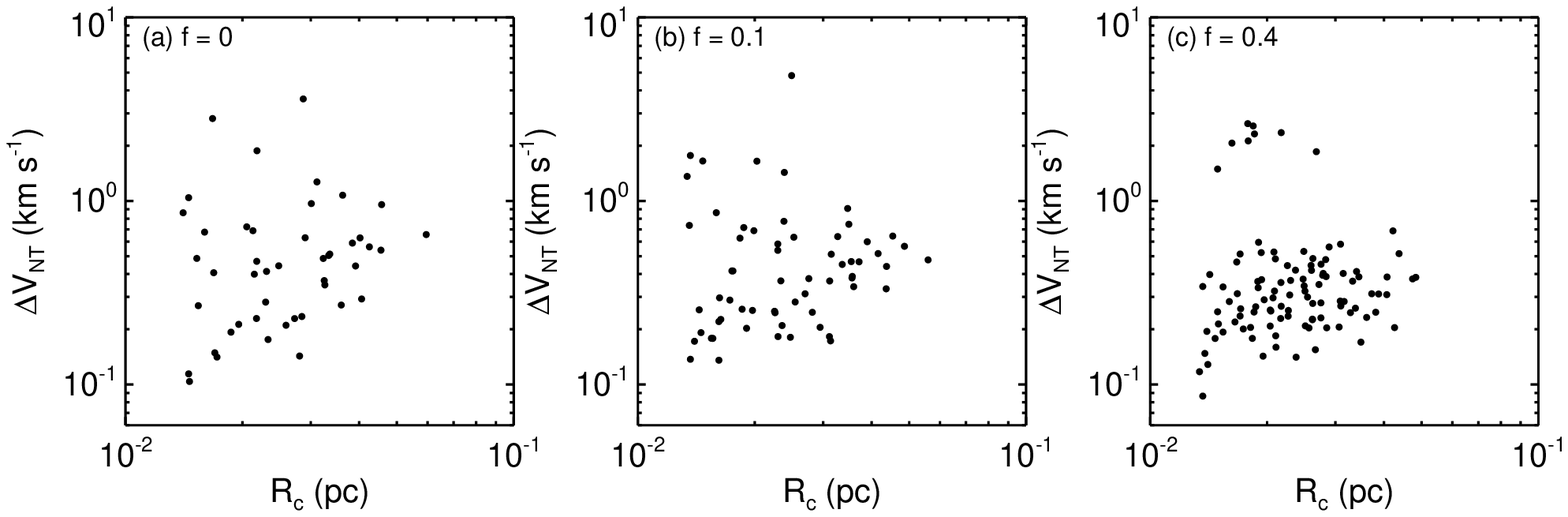}
\caption{Same as Fig. \ref{fig:velocity-radius} but for 
models (a) S3, (b) S2, and (c) S1. 
}  
\label{fig:outflow}
\end{figure}

In Figs. \ref{fig:outflow}a through \ref{fig:outflow}c,  
we compare the velocity dispersion-radius relations
for the moderately-strong magnetic models with no outflow feedback
($f=0$), weak outflow $(f=0.1)$, and strong outflow 
($f=0.4$).
In the model with no outflow feedback (model S3), the number of identified 
cores is smaller than in the other two models, and the mean core mass 
tends to be slightly larger.  This difference may be due to the fact that 
the ambient turbulent motions are weaker without any outflow feedback 
and thus the turbulent fragmentation does not form smaller-scale 
structures as efficiently.  
Figure \ref{fig:outflow} indicates that the velocity dispersion-radius
relation does not depend strongly on the strength of the outflow feedback.
For all three models, there is no clear relationship between 
the velocity dispersion and the core radius.
Thus, it appears difficult to discriminate the effects of 
protostellar outflows based on the velocity dispersion-radius 
relation alone.

\begin{figure}
%\epsscale{0.8}
\plotone{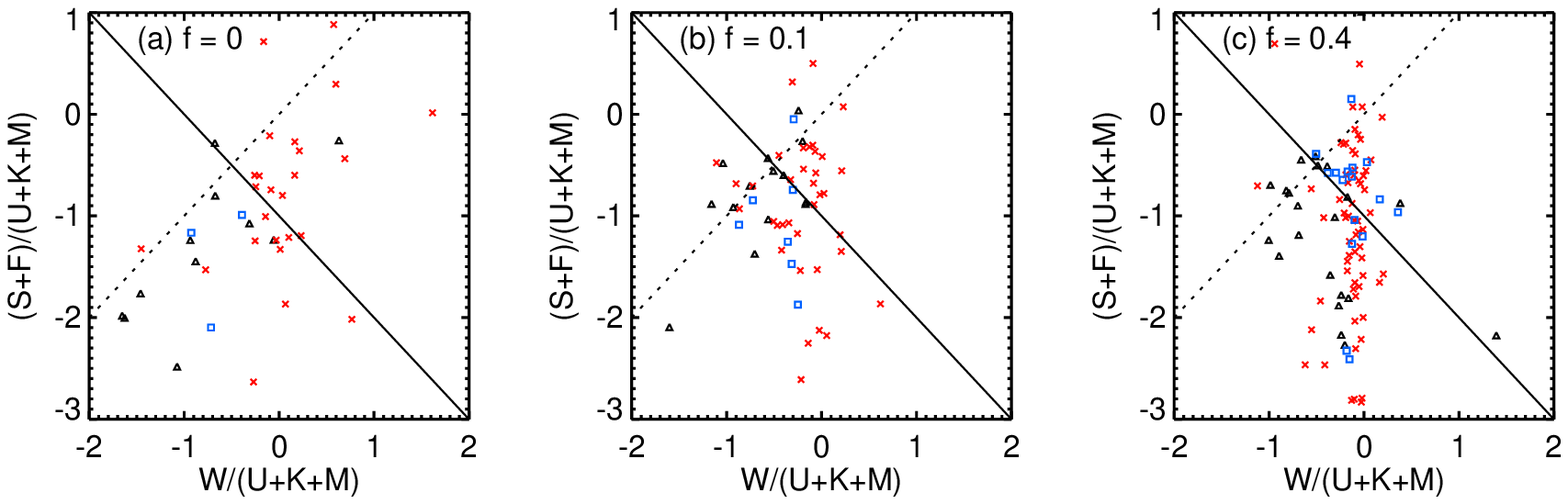}
\caption{Same as Fig. \ref{fig:virial analysis} but for 
models (a) S3, (b) S2, and (c) S1.
}  
\label{fig:outflow2}
\end{figure}

The surface term-gravitational term plots in Figs. \ref{fig:outflow2}a 
through \ref{fig:outflow2}c indicate that the dynamical states of 
the cores depend somewhat on the outflow strength.
In the absence of the outflow feedback, more cores are located near 
the dashed line (where the two terms are equal). The gravitational 
term appears more important for this model than for the other two, 
although the surface term is clearly still important. This is due,
at least in part, to the fact that the global gravitational infall 
can generate fast motions even in the absence of protostellar
outflow driven turbulence.

\begin{figure}
%\epsscale{0.8}
\plotone{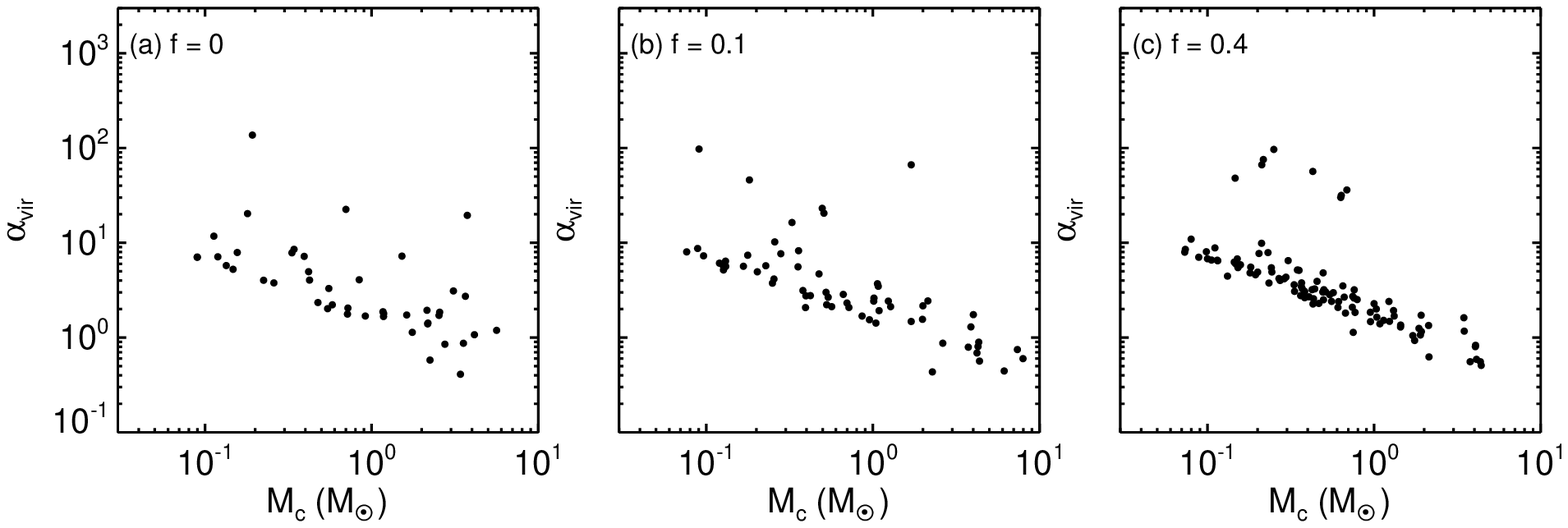}
\caption{Same as Fig. \ref{fig:virial parameter} but for 
models (a) S3, (b) S2, and (c) S1.
}  
\label{fig:outflow3}
\end{figure}

It is difficult, however, to gauge the effects of the outflow feedback 
from the virial parameters of the cores, which are often 
used in observational studies of the molecular cloud cores in 
star-forming molecular clouds. The difficulty is illustrated in 
Figs. \ref{fig:outflow3}a through \ref{fig:outflow3}c, where we show
the virial parameter-mass relations for the three models
with different outflow strengths.
In all cases, the distributions of the virial parameters are similar, 
following a power-law of $\alpha_{\rm vir} \propto M_{\rm c}^{-2/3}$
with rather small dispersions.

\subsection{Outflow Direction}

In this subsection, we investigate how well the outflow axes are aligned 
with the global or initial magnetic field direction in parsec-scale
cluster forming clumps.
In our simulations, the outflow direction is set to
the local magnetic field direction at the position of each formed 
star. This should be a reasonable  
approximation since the outflows are most likely driven along the
local magnetic field by field line twisting due to disk rotation,
\citep[e.g.,][]{matsumoto06}.

\begin{figure}
%\epsscale{0.8}
\plotone{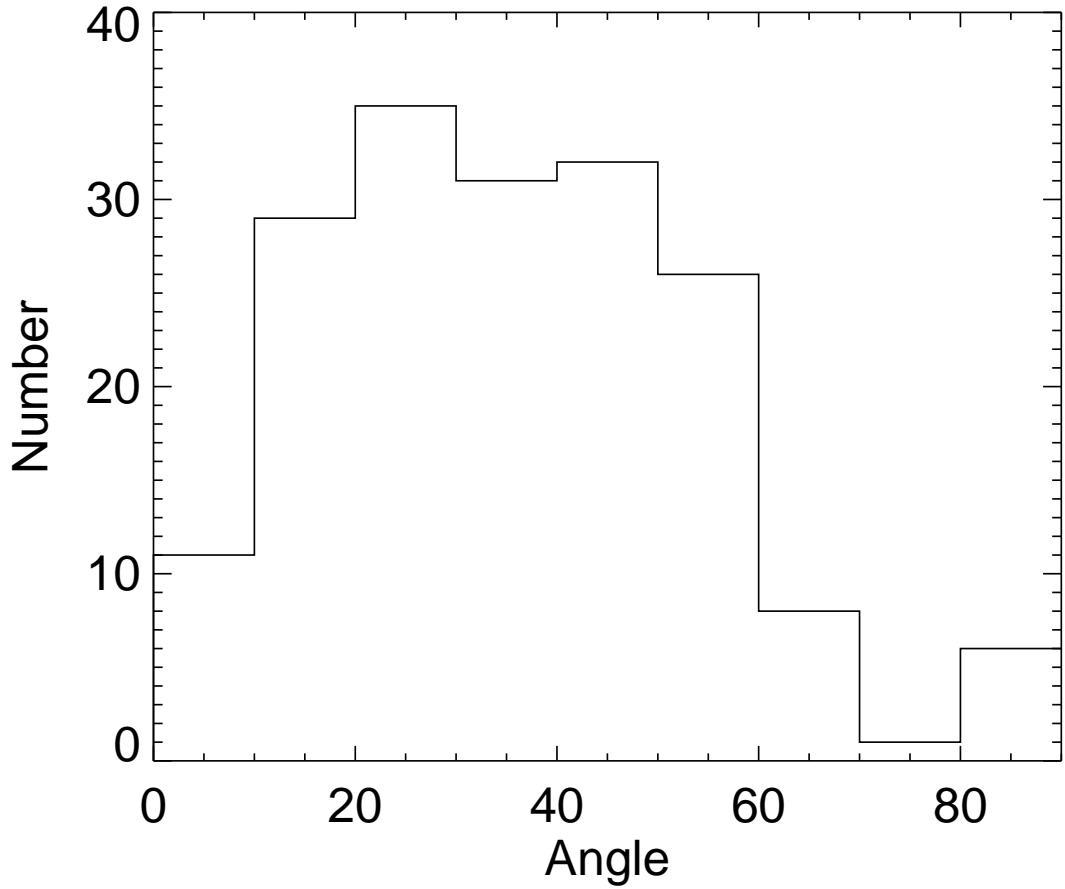}
\caption{Histogram of the outflow direction relative to the initial
 magnetic field direction for the model with a moderately-strong
 magnetic field (model S1).
}  
\label{fig:outflow4}
\end{figure}

Figure \ref{fig:outflow4} shows the histogram of the outflow 
direction as a function of the angle between the outflow axis 
and the initial magnetic field direction, for the model with 
a moderately-strong magnetic field (model S1).
All the outflows created by the time when $SFE \simeq 0.16$ are 
shown.
The outflow direction shows a broad distribution, taking its maximum
at the mean angle of about 30$^\circ$. The conclusion is
that, although there is some preference for the outflows to align
with the global magnetic field direction, the alignment is not as
strong as one may naively expect. This is because most of the dense cores 
are formed through outflow-induced turbulent compression, which can 
push the magnetic field on the core scale away from the global field 
direction. On smaller scales, the misalignment in field direction can 
be further amplified by gravitational collapse. The lack of a 
complete alignment between the outflow axes and the global field
direction is thus expected. It can not be used to argue against 
the presence of a moderately-strong global magnetic field.

\section{Discussion}
\label{sec:discussion}

\subsection{Comparison with Core Observations}
\label{subsec:observations}

Here, we compare the core properties obtained in Section 
\ref{sec:results} with observations.
We use the core data toward two nearby parsec-scale 
cluster-forming clumps: 
the $\rho$ Ophiuchus Main Cloud and NGC 1333 in Perseus. 
In the  $\rho$ Ophiuchus region, \citet{maruta10} 
identified 68 H$^{13}$CO$^+$ ($J=1-0$) cores by applying clumpfind 
to the 3D position-position-velocity cube data obtained using 
Nobeyama 45~m telescope. In NGC~1333, \citet{walsh07} identified 
93 N$_2$H$^+$ ($J=1-0$) dense cores, again using clumpfind.
Since both the H$^{13}$CO$^+$ ($J=1-0$) and  N$_2$H$^+$ ($J=1-0$)
transitions have critical densities of order $\sim 10^5$ cm$^{-3}$, 
the H$^{13}$CO$^+$ and N$_2$H$^+$ cores should have densities 
comparable to our simulated cores, which were identified using 
a threshold density $\sim 10^5$ cm$^{-3}$. Since the moderately-strong 
field model yields a core velocity dispersion that is in better 
general agreement with observations than the non-magnetic or 
weak-field model, we will concentrate on comparing its cores 
with the $\rho$ Oph and NGC~1333 data. 

\begin{figure}
%\epsscale{0.8}
\plotone{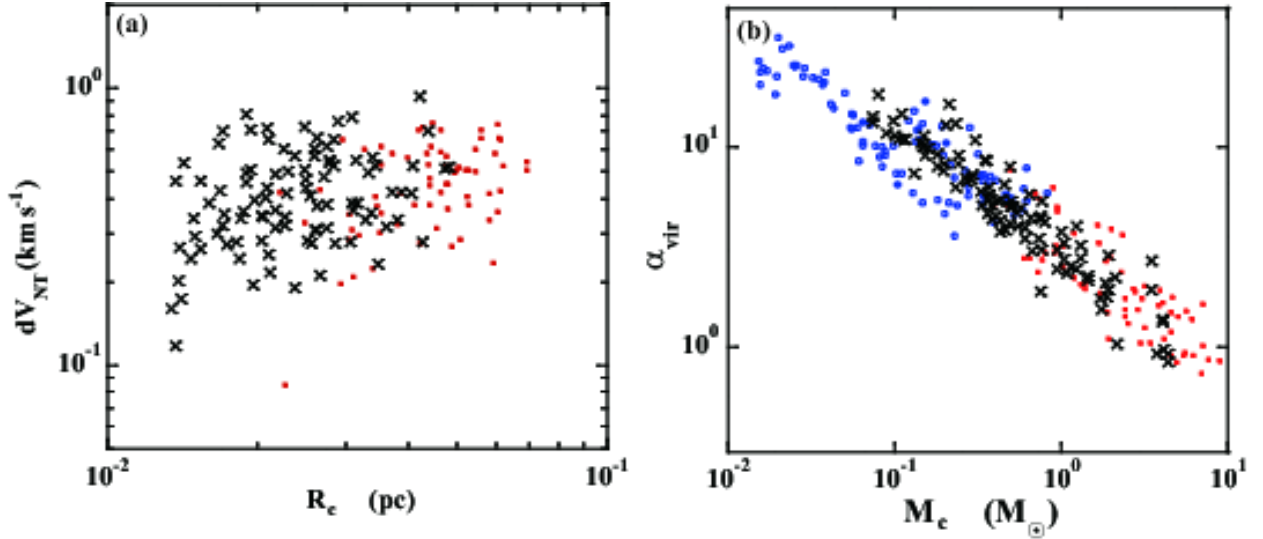}
\caption{(a)
the relationship between the 1D FWHM nonthermal velocity width and 
core radius for the dense cores in the $\rho$ Ophiuchi Main Cloud
and our cores identified from the moderately-strong magnetic field model
(model S1).
The red dots and black crosses denote the $\rho$ Oph cores and our
cores of model S1, respectively.
For the $\rho$ Oph cores, we used the results of \citet{maruta10} who 
identified the cores using H$^{13}$CO$^+$ ($J=1-0$) data.
For our cores of model S1, we recalculated the 1D FWHM velocity widths from 
the 3D core velocity dispersions derived in Section \ref{sec:results}.
(b) The virial parameter-core mass relation.  
The red dots and black crosses are the same as those of panel (a).
The blue open circles denote the cores of NGC 1333 identified using
N$_2$H$^+$ ($J=1-0$) emission \citep{walsh07}. 
For $\rho$ Oph cores and NGC 1333 cores, the LTE masses are adopted as
the core masses.
For the NGC 1333 cores, we recalculated the LTE masses 
derived by \citep{walsh07} 
by assuming the average N$_2$H$^+$ fractional abundance 
relative to H$_2$ obtained toward $\rho$ Oph of $X_{\rm N_2H^+} \sim 
3\times 10^{-10}$ and
distance of 250 pc. See the text for detail.
}  
\label{fig:obs}
\end{figure}

In Fig. \ref{fig:obs}a, we compare the 1D FWHM nonthermal velocity
width-radius relations for the $\rho$ Oph cores of \citet{maruta10} 
and our simulated cores (see Section \ref{sec:results}). The 
3D velocity dispersions discussed in Section \ref{sec:results} 
was converted into the 1D FWHM velocity widths using $dV_{\rm NT} 
= (8\ln2/3)^{1/2}\Delta V_{\rm NT}$. Our simulated cores tend 
to be somewhat smaller than the $\rho$ Oph cores; the average 
radius of the $\rho$ Oph cores is estimated at 0.04 pc, about 
1.7 times that of model S1. Aside from the size difference, the 
observed and simulated data are almost indistinguishable. Neither shows
a clear dependence of linewidth with size, except for a weak  
trend for the linewidth to increase slightly with radius. Neither
follows the Larson's linewidth-size relation. The observations are 
also consistent with the cores formed from initially-magnetically 
subcritical clouds \citep[see Fig. 18 of ][]{nakamura08}. We did 
not plot the NGC 1333 cores on Fig. \ref{fig:obs}a because the 
core radius in Table 1 of \citet{walsh07} was quoted in single 
significant digit, which artificially introduces a large scatter 
in the plot. Nevertheless, inspection by eye reveals no clear 
correlation between the velocity width and core radius,   
broadly consistent with the data for both the $\rho$ Oph cores 
and the simulated cores of model S1.

Figure \ref{fig:obs}b compares the virial parameter-mass relations of
the simulated and observed cores.
The red and blue dots, and black crosses denotes the $\rho$ Oph  
\citep{maruta10} and NGC 1333 cores \citep{walsh07}, and the cores of model S1, respectively.
For the NGC 1333 cores, we recalculated the virial masses by including
the thermal contribution from the gas with a temperature of 15 K and  
adopting a distance of 235 pc
\citep{hirota08}, instead of 300 pc used in \citet{walsh07}.
We also recalculated the LTE masses by adopting the N$_2$H$^+$
fractional abundance relative to H$_2$ of $X_{\rm N_2H^+} \sim 
3\times 10^{-10}$ derived from observations 
toward $\rho$ Oph 
because \citet{walsh07} determined their adopted N$_2$H$^+$ 
fractional abundance such that the points are distributed 
evenly about the line of equality ($M_{\rm LTE}=M_{\rm vir}$)
in their $M_{\rm LTE}$-$M_{\rm vir}$ plot.
This N$_2$H$^+$ fractional abundance is determined by taking 
average of the values obtained toward Oph A \citep{james04} 
and B2 subclumps \citep{friesen10}.
Figure \ref{fig:obs}b indicates that 
the virial parameter-core mass relation of the nearby cluster-forming
clumps follows roughly a single power-law of 
$\alpha_{\rm vir} \propto M^{-2/3}_{\rm c}$ 
with a small dispersion, which is again in good agreement with model  
S1 that has a moderately-strong magnetic field. A dynamically 
important magnetic field may have played a role in core formation 
in these regions.

Besides the $\rho$ Oph region, \citet{maruta10} found that the velocity 
dispersions of H$^{13}$CO$^+$ cores are almost independent of the core 
radius in Orion A as well, even though the latter is more than a factor 
of 3 farther away. The coarser spatial resolution for Orion means that its 
cores may contain (smaller, $\rho$ Oph-like) cores blending together. 
If this is true, then 
the nearly flat velocity-width-radius relation observed in both regions  
suggests that the inter-core motions among the neighboring cores 
are almost comparable to the internal motions in the 
individual cores. Such a feature was pointed out by \citet{andre07} 
who measured the velocity difference among neighboring cores using 
N$_2$H$^+$ ($J=1-0$). 
If the velocity-width-radius relation is flat 
($\Delta V \propto$ constant) and the core mass is proportional to 
$R_{\rm c}^3$, then the virial parameter is scaled 
as $\alpha_{\rm vir} \propto R_{\rm c} \Delta V^2 / (GM_{\rm c})
\propto M_{\rm c}^{-2/3}$, a similar power-law to that 
derived by \citet{bertoldi92}.
This again suggests the importance of ambient turbulent
pressure in dynamics of the cores as discussed 
in Section \ref{sec:results}.

\subsection{Amplification of Magnetic Field}
\label{subsec:amplification}

\begin{figure}
%\epsscale{0.8}
\plotone{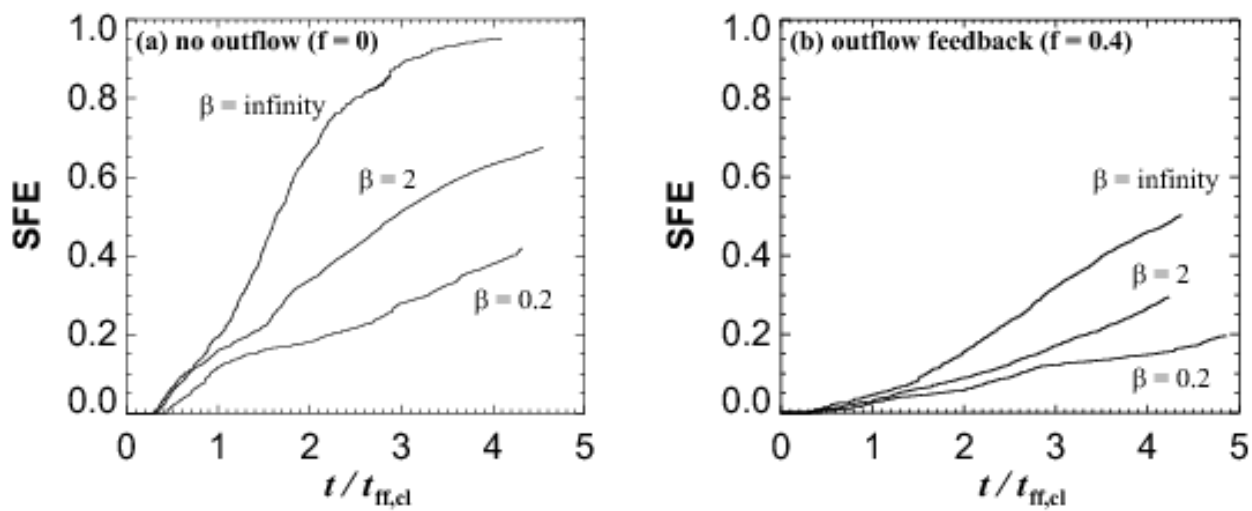}
\caption{(a) Time evolution of star formation efficiencies for 
three models with no outflow feedback. The three models have
different initial magnetic field strengths: models N2
($\beta=\infty$), W2 ($\beta = 2$), and S2 ($\beta = 0.2$).
The evolution time is normalized to the global clump free-fall time,
$t_{\rm ff,cl}=0.49 (L/1.5 {\rm pc})(20 {\rm K}/T)^{1/2}$ Myr.  
(b) Same as panel (a) but for models with outflow feedback
(models N1, W1, and S1).
}  
\label{fig:sfe}
\end{figure}

To isolate the effect of magnetic field on clustered star formation,  
we present in Fig. \ref{fig:sfe}a the time evolution of star 
formation efficiency (SFE) for three models with different field strengths 
but no outflow feedback. Our simulations show that, as expected, the 
magnetic field tends to slow down star formation, especially at
relatively late times. The rate of star formation remains rather 
high, however, particularly at early times. For example, from the 
formation of the first star to the time when the SFE reaches $\sim 
16 \%$, the average star formation efficiency per global free-fall 
time (SFR$_{\rm ff}$; Krumholz \& McKee 2005)
is estimated  to be 
29 \%, 27 \%, and 21 \%, respectively, for models N2, W2, and S2. 
These values are larger than the observationally inferred values 
of a few \% \citep{krumholz07}.
The large SFR$_{\rm ff}$ in the absence of outflow feedback is in agreement 
with previous studies (see also the Appendix for the analytic formula
of SFR$_{\rm ff}$). 
For example, using MHD SPH simulations, 
\citet{price08} and \citet{price09} followed the
evolution of 50 $M_\odot$ clumps 
with $0.25 \le \beta \le  \infty$ until $t\lesssim 1.5 t_{\rm ff,cl}$.
Although their initial model clump mass is too small compared to 
nearby cluster-forming clumps like $\rho$ Oph and NGC 1333 (which 
have masses of order $10^3$ M$_\odot$), their star formation
efficiencies per free-fall time are estimated to be over 
10 \%.  These simulations imply that it is difficult to reduce the 
star formation rate to the observed level by a moderately-strong 
magnetic field alone, and other factors are needed to significantly 
retard star formation in cluster-forming clumps.

One way to slow down star formation is the feedback from forming stars.
\citet{price09} found that radiative feedback from forming 
stars does not change 
the global star formation efficiency in a parent clump much, although  
it significantly suppresses the small-scale fragmentation by increasing the
temperature in the high-density material near the protostars.
Figure \ref{fig:sfe}b shows that the star formation rate is greatly reduced 
by the inclusion of outflow feedback, even in the presence of a
relatively weak magnetic field, in agreement with previous studies 
\citep{li06,nakamura07,wang10}. The average star formation 
efficiency per global free-fall time is estimated to be 8 \%, 6 \%, 
and 4 \% for model N1 (no magnetic field), W1 (weak field), and 
S1 (moderately-strong field), respectively. For the model with the 
moderately-strong field, the SFR$_{\rm ff}$ is more or less comparable to the 
observed value.

\begin{figure}
%\epsscale{0.8}
\plotone{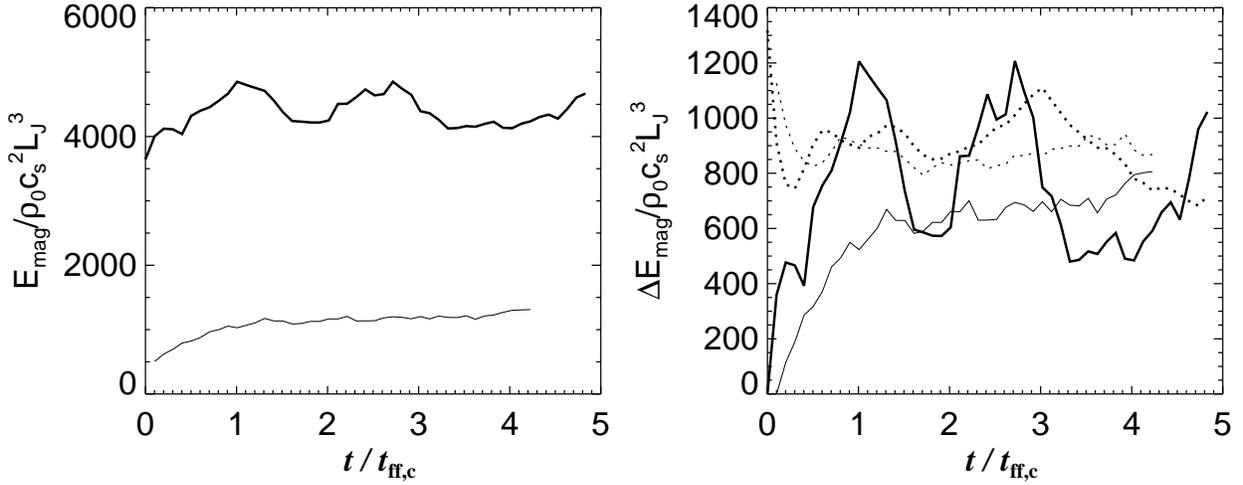}
\caption{(a) Time evolution of total magnetic energy for two models with
different initial magnetic field strengths. Thick and thin lines are 
for the models with moderately-strong (model S1) and weak magnetic field
(model W1), respectively. 
The evolution time is normalized to the global clump free-fall time,
$t_{\rm ff,cl}=0.49 (L/1.5 {\rm pc})(20 {\rm K}/T)^{1/2}$ Myr.  
(b) Time evolution of magnetic energy of
amplified component (solid lines) and kinetic energy of turbulence 
(dotted lines). The magnetic energy of the amplified component tends to 
approach the kinetic energy for each model.}  
\label{fig:magnetic field}
\end{figure}

\begin{figure}
\epsscale{0.8}
\plotone{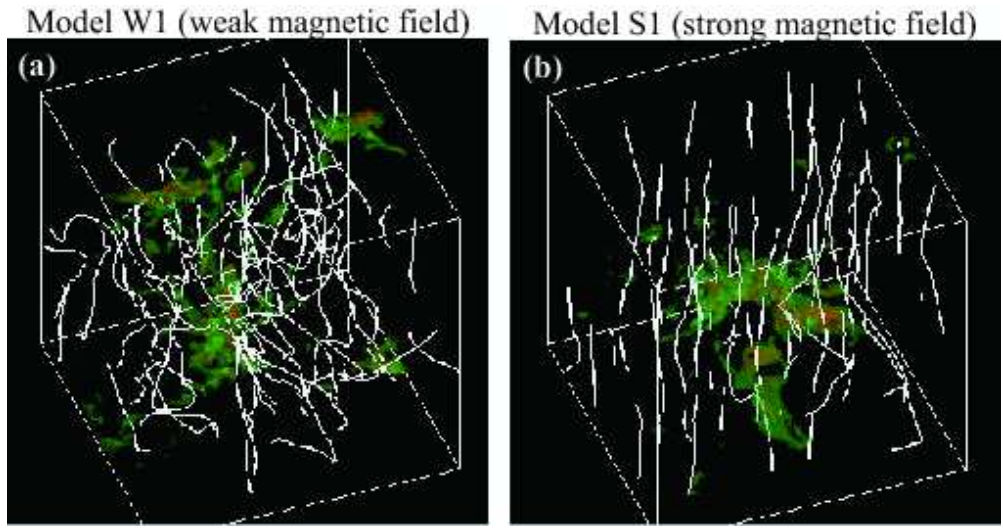}
\caption{3D view of the density and magnetic field distributions for 
(a) the weakly-magnetized model (model W1) 
and (b) the strongly-magnetized model
(model S1) at the stage where the star formation efficiency has reached 
0.16. The color images indicate the isodensity surfaces.
The white curves are the magnetic field lines.
For both the panels, only the central $L/2 \times L/2 \times L/2$ region 
is shown.
}  
\label{fig:3d}
\end{figure}

The reason for the large reduction in SFR$_{\rm ff}$ can be 
seen in Fig. \ref{fig:magnetic field}, where the magnetic 
energy is plotted against the evolution time, for models W1 and S1.
As shown in Fig. \ref{fig:magnetic field}a, for the moderately-strong 
field case,  
the total magnetic energy is dominated by the background uniform 
field that does not contribute to the force balance in the initial
cloud.  We therefore illustrate in Fig. \ref{fig:magnetic field}b
the time evolution of the magnetic energy stored in the 
distorted component that was amplified by supersonic turbulence. 
Here, we computed the magnetic energy stored in the distorted component
by subtracting the initial magnetic energy from the total magnetic
energy.
For comparison, we also plotted in Fig. \ref{fig:magnetic field}b 
the evolution of 
the total kinetic energy of the clump. Since the kinetic energy is often dominated by the
unbound high-velocity gas associated with the active outflows, 
we approximate the total kinetic energy of the clump as the kinetic
energy of the gas whose velocity is smaller than 10 $c_s$.
For both models, the amplified components tend to increase with time.
For the model with the moderately-strong field,  
the amplified component begins to oscillate around a level value 
after a free-fall time. For both models, the magnetic energy of 
the amplified component becomes comparable to the kinetic energy 
of the dense gas by the end of the computation, indicating that 
a quasi-equipartition has been reached 
\citep[see also ][]{federrath11}.
For the weak field case, the amplified component is more 
important than the initial uniform field, resulting in 
a significantly-distorted magnetic field structure.
Such a highly-distorted magnetic field can be seen in the 3D bird's eye
view of the  density and magnetic field distributions of model W1,
which is presented in the left panel of Fig. \ref{fig:3d}.
In contrast, the global magnetic field is well-ordered for the
case of moderately-strong initial magnetic field, which is shown in
the right panel of Fig. \ref{fig:3d}.

\subsection{Polarization Maps}
\label{subsec:polarization}

\begin{figure}
%\epsscale{0.8}
\plotone{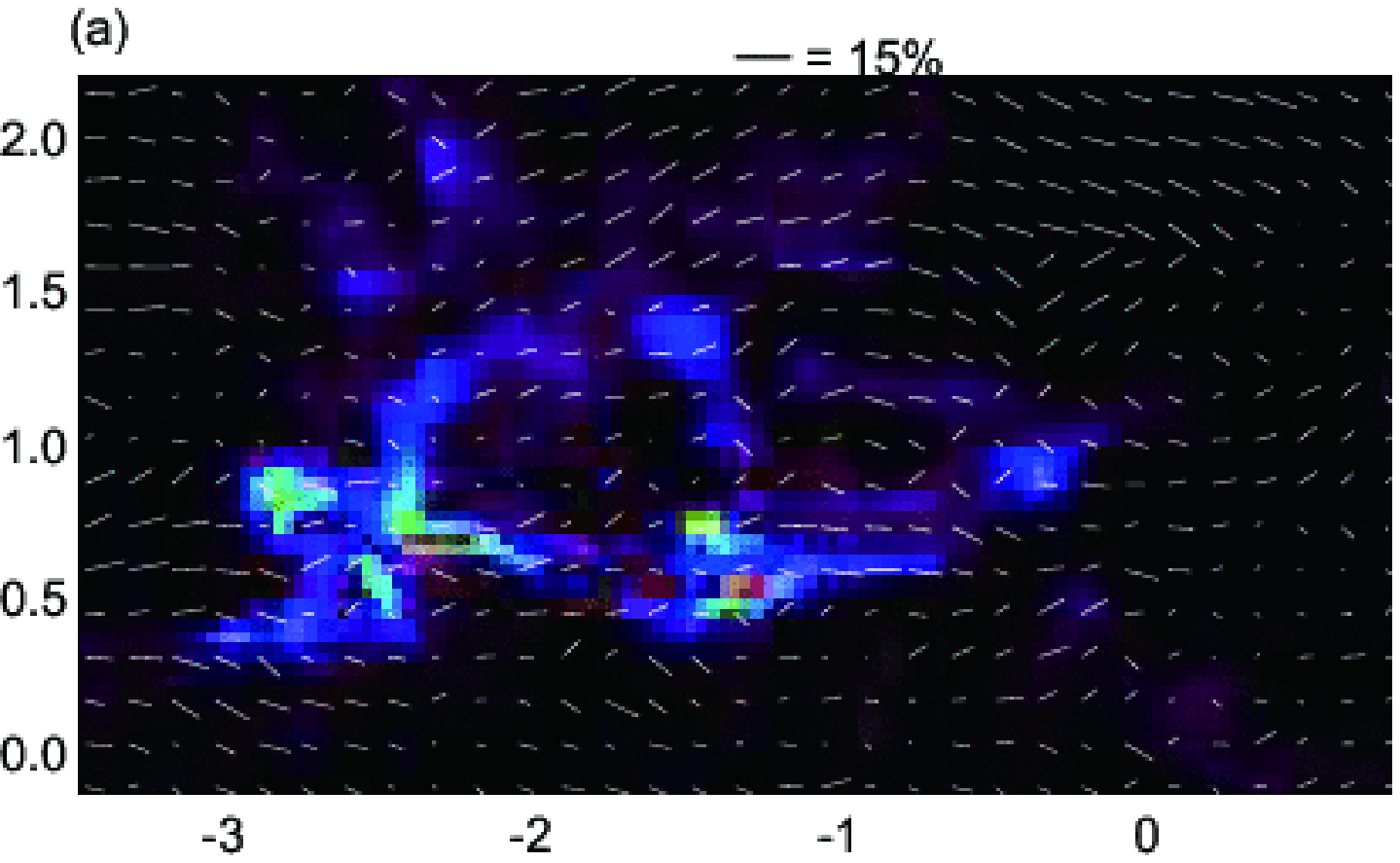}
\plotone{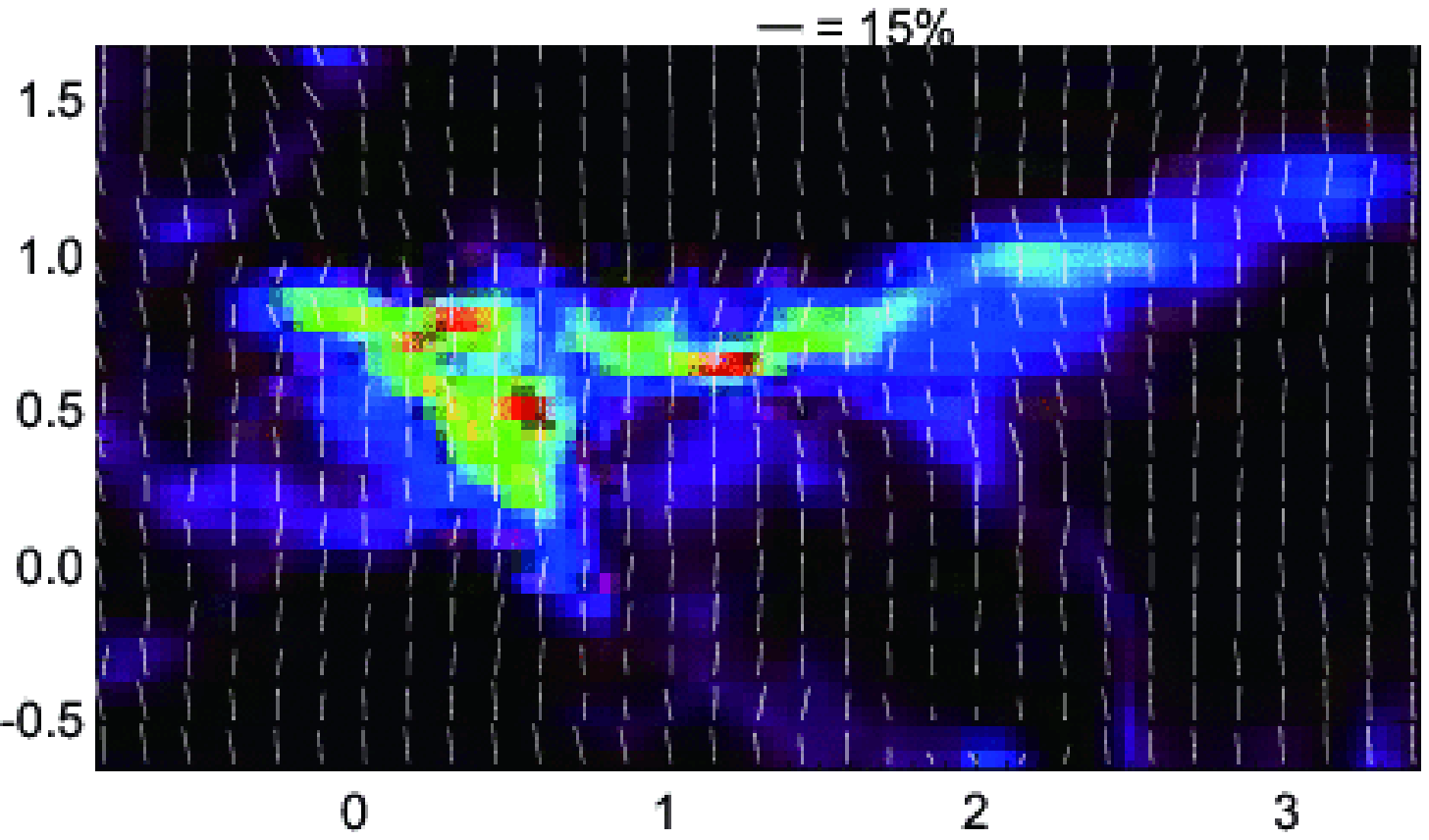}
\caption{(a) Polarization maps of (a) model W1
(weak magnetic field) and (b) model S1 
(moderately-strong magnetic field). 
The length of a polarization vector is proportional to the degree
of polarization, with the longest vector corresponding to $P=15\%$.
Only one polarization vector is plotted for every four computational 
cells. The color contour shows the column density distribution. The initial
magnetic field lines are parallel and perpendicular to the horizontal
line for panels (a) and (b), respectively.
The units of length are the central Jeans length
$L_J=(\pi c_s^2/G\rho_0)^{1/2} \simeq 0.17 
(T/20 {\rm K})^{1/2}(n_{\rm H_2,0}/2.69\times
 10^4 {\rm cm}^{-3})^{-1/2}$ pc.}  
\label{fig:polarization}
\end{figure}

Polarization maps of submillimeter thermal dust emission have recently 
been obtained for nearby star forming regions
\citep[e.g.,][]{houde04,girart09,matthews09}.
Here, we present the polarization maps derived from the simulation data 
for the two magnetically-supercritical models 
with different initial magnetic field strengths (models W1 and S1
which have the magnetic flux-to-mass ratios of 
$\bar{\lambda}=4.3$ and 1.4, respectively).
We computed the polarized thermal dust emission from the MHD model
following \citet{padoan01}. We neglect the effect of self-absorption and 
scattering because we are interested in the thermal dust emission at
submillimeter wavelengths.  We further assume that the grain properties
are constant and the temperature is uniform.
The polarization degree is set such that the maximum is equal to 15 \%.
Figure \ref{fig:polarization} shows the dust polarization maps
calculated from models W1 and S1.

Only a small portion of the computation box is shown in each panel
of Fig. \ref{fig:polarization}.
As expected from Fig. \ref{fig:3d}, in the presence of a weak 
magnetic field, the spatial distribution of the polarization vectors 
has relatively large fluctuations.  
The polarization degree tends to
be smaller in several parts 
where the magnetic fields are strongly distorted.
In the densest parts, the polarization vectors appear 
more or less parallel to the local elongated structures or 
dense filaments, whereas there is no clear correlation between the 
vectors and density distribution in less dense parts.
This may be due to the fact that the local dense filaments are 
created by turbulent compression that preferentially enhances the 
magnetic field component transverse to the shock plane.
In contrast, in the presence of the strong magnetic field, 
the filamentary structure is prominent and the filament axes tend to 
be perpendicular to the polarization vectors that are almost parallel 
to the initial magnetic field direction, suggesting that 
the gas flow is channeled preferentially 
by the strong magnetic field to form the filaments.
The polarization observations 
can thus provide a handle on the magnetic field strength of cluster-forming 
clumps.

Recent polarization observations of cluster-forming
clumps show that the global magnetic field lines are more or less spatially 
well-ordered \citep[e.g.,][]{davis00,houde04,sugitani10,sugitani11}. 
For example, \citet{sugitani10} found that the 
Serpens cloud core is threaded by a hour-glass shaped well-ordered
magnetic field and is elongated in the cross-field direction. 
\citet{sugitani11} found that the Serpens 
South filamentary infrared dark cloud appears to be threaded 
by more or less straight global magnetic field. 
The observed spatially-ordered magnetic field structures imply 
that the magnetic fields in the nearby cluster-forming
clumps are likely to be at least moderately-strong and dynamically 
important.

\subsection{Core Mass Function}
\label{subsec:CMF}

\begin{figure}
%\epsscale{0.8}
\plotone{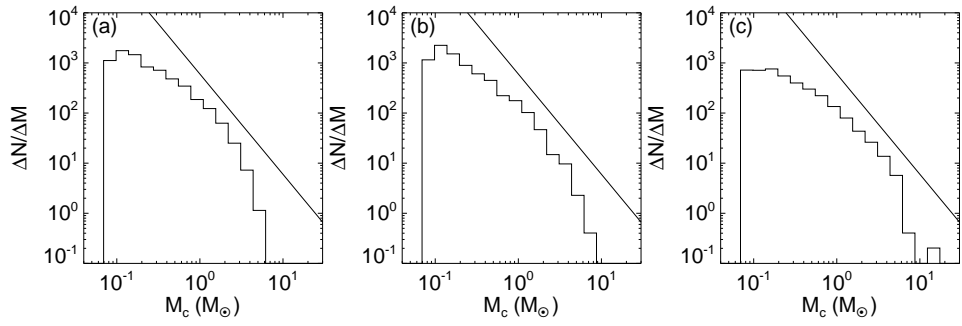}
\caption{Core mass functions for (a) model N1, (b) model W1 and (c)
model S1. A power-law of $\Delta N/\Delta M \propto M^{-2}$ is plotted
 for comparison.  
}  
\label{fig:CMF}
\end{figure}

In Figs. \ref{fig:CMF}a, \ref{fig:CMF}b, and \ref{fig:CMF}c, 
we plot the core mass functions (CMFs) for our identified cores for models 
N1, W1, and S1, respectively. 
To enlarge the number of sample cores, we added up all the cores 
identified at the three stages when SFE has reached 0.08, 0.12, and 0.16.
In addition, we added up the cores identified from the runs 
using the different initial turbulent realization 
for which the different random numbers were used 
for both the amplitude and phase.
The total number of cores so identified is over 500 for each model.

The overall shapes of the CMFs are similar in all three cases,
suggesting the shape of the CMF is relatively insensitive 
to the initial magnetic field strength. 
This is different from the 
turbulent fragmentation scenario proposed by \citet{padoan01} 
where a weak magnetic field is required to reproduce the CMF
that is similar in shape to the Salpeter IMF
\citep[see also][]{li10}. In our simulations, 
a moderately-strong magnetic field can produce a CMF that resembles the 
Salpeter IMF as well.  

We note two interesting features of the computed CMFs. 
First, there appear to be a lack of 
massive cores compared to the Salpeter stellar IMF. This is not 
necessarily a problem, because it is unclear whether the nearby
cluster-forming clumps such as $\rho$ Oph and NGC 1333 that we
aim to simulate would ever produce massive stars. Alternatively, 
massive stars can grow from initially less massive cores, fed by 
global collapsing flows towards the bottom of the clump potential 
(Smith et al. 2008; Wang et al. 2010). Second, the turnover at 
the low mass end is less clear in the CMFs than in the Kroupa or 
Chabrier IMF. This may be because, at the low mass end, most of 
the cores are gravitationally unbound, and may not form stars. 
Further studies are required to determine the relation between 
the CMF and IMF.

\section{Summary}
\label{sec:summary}

We have performed a set of 3D MHD simulations of cluster formation taking into
account the effects of protostellar outflow feedback, and identified 
dense cores by applying a clumpfind algorithm to the simulated 3D 
density data cubes. The main results are as follows.

1. Dense cores do not follow Larson's linewidth-size relation. We find
that the velocity dispersions of dense cores show little correlation 
with core radius, irrespective of the strength of the magnetic field 
and outflow feedback. In the absence of a magnetic field, the 
majority of the cores have supersonic velocity dispersions, whereas 
in the presence of a moderately-strong magnetic field, the cores 
tend to be subsonic or at most transonic. 

2. We find that most of the cores are out of virial equilibrium, with 
the external pressure due to ambient turbulence dominating the 
self-gravity. The core formation and evolution is largely controlled
by the dynamical compression due to outflow-driven turbulence.
Such a situation is contrast to the strongly-magnetized (magnetically
subcritical) case, where the self-gravity plays a more important 
role in the core dynamics, particularly for massive cores.

3. Even an initially-weak magnetic field can retard star formation 
significantly, because the field is amplified by supersonic 
turbulence to an equipartition strength.  
In such an initially weak field, the distorted field component 
dominates the uniform one. In contrast, for a moderately-strong field, 
the uniform component remains dominant. Such a difference in the 
magnetic structure can be observed in simulated polarization maps 
of dust thermal emission. Recent polarization measurements show 
that the field lines in nearby cluster-forming clumps are 
spatially well-ordered, indicative of a moderately-strong,
dynamically-important, field.

\acknowledgments 
This work was supported in part by a Grant-in-Aid for Scientific Research
of Japan (20540228 and 22340040) and NASA (NNG06GJ33G and NNX10AH30G). 
The numerical calculations were carried out mainly on
NEC SX9 at the Center for Computational Astrophysics 
(CfCA) at National Astronomical Observatory of Japan 
and on NEC SX8 at YITP in Kyoto University.

\appendix

\section{Star Formation Rate in A Cluster Forming Clump}

Using 3D MHD numerical simulations of cluster formation,
\citet{li06} and \citet{nakamura07} demonstrated that 
the protostellar outflow-driven turbulence can keep 
a pc-scale, cluster-forming clump close to a virial equilibrium long
after the initial turbulence has decayed away.
Here, we derive an analytic formula of star formation rate
in a cluster-forming clump that keeps its virial equilibrium 
by the protostellar outflow feedback.

Numerical simulations of protostellar turbulence indicate that 
the dissipation rate of the turbulence momentum, 
$dP_{\rm turb}/dt$, balances 
the momentum injection rate by the protostellar outflow feedback, 
$dP_{\rm out}/dt$, so that the clump can be kept close to a virial equilibrium, 
\begin{equation}
\frac{dP_{\rm turb}}{dt} = \frac{dP_{\rm out}}{dt} \ .
\label{eq:turbulence}
\end{equation}

The dissipation rate of the turbulence momentum, $dP_{\rm turb}/dt$, 
can be written as
\begin{equation}
\frac{dP_{\rm turb}}{dt} = \alpha \frac{M_{\rm cl} V_{\rm vir}}{t_{\rm diss}}
\ , 
\end{equation}
where $\alpha \sim 0.5 $ \citep{maclow99}. 
The virial velocity, $V_{\rm vir}$, is given by
\begin{equation}
V_{\rm vir} = \sqrt{\frac{3}{5}af_{\rm B}\frac{GM_{\rm cl}}{R_{\rm cl}}}
 \ . 
\end{equation}
The momentum dissipation time is 
$t_{\rm diss} = R_{\rm cl}/V_{\rm vir}$.
The dimensionless parameter of order unity, $a$, measures the
effects of a nonuniform or nonspherical mass distribution.
For a uniform sphere and a centrally-condense sphere with
$\rho \propto r^{-2}$, $a=1$  and 5/3, respectively.
In the following analysis, we adopt $a=5/3$ because 
the cluster-forming clumps tend to be centrally-condensed. 
We take into account the magnetic support by multiplying 
the virial velocity by a factor $f_{\rm B}$, where
$0 \lesssim f_{\rm B} \lesssim 1$.

The momentum injection rate by the protostellar outflows 
is given by
\begin{equation}
\frac{dP_{\rm out}}{dt} = \epsilon_{\rm SFR} \times f_{\rm w} V_{\rm w} \ ,
\label{eq:outflow}
\end{equation}
where $V_{\rm w}$ is the flow speed and 
$\epsilon _{\rm SFR}$ is the star formation rate.
Using equations (\ref{eq:turbulence}) and (\ref{eq:outflow}),
the star formation rate is rewritten as
\begin{equation}
\epsilon_{\rm SFR} = \frac{3}{5}a \alpha f_{\rm B} f_{\rm w}^{-1}V_{\rm w}^{-1} 
\frac{GM_{\rm cl}^2}{R_{\rm cl}^2} \ .
\end{equation}
If we normalize $\epsilon_{\rm SFR}$ to $M_{\rm cl}/t_{\rm ff}$,
then 
the star formation efficiency per free-fall time
is given by
\begin{equation}
{\rm SFR}_{\rm ff} = \epsilon _{\rm SFR} \frac{t_{\rm ff}}{M_{\rm cl}}
\simeq 0.0125 \alpha \left(\frac{f_{\rm B}}{0.5}\right) 
\left(\frac{f_{\rm w}}{0.5}\right)^{-1}
\left(\frac{V_{\rm w}}{10^2 \ {\rm km \ s^{-1}}}\right)^{-1}
\left(\frac{R_{\rm cl}}{1\ {\rm pc}}\right)^{1/2}
\left(\frac{\Sigma_{\rm cl}}{5 \times 10^{21} {\rm
cm^{-2}}}\right)^{1/2} \ ,
\end{equation}
where the surface density is $\Sigma = M_{\rm cl}/\pi R_{\rm cl}^2$  and
the free-fall time is $t_{\rm ff}=\sqrt{3\pi/32G\rho}$.

\begin{figure}
\epsscale{0.8}
\plotone{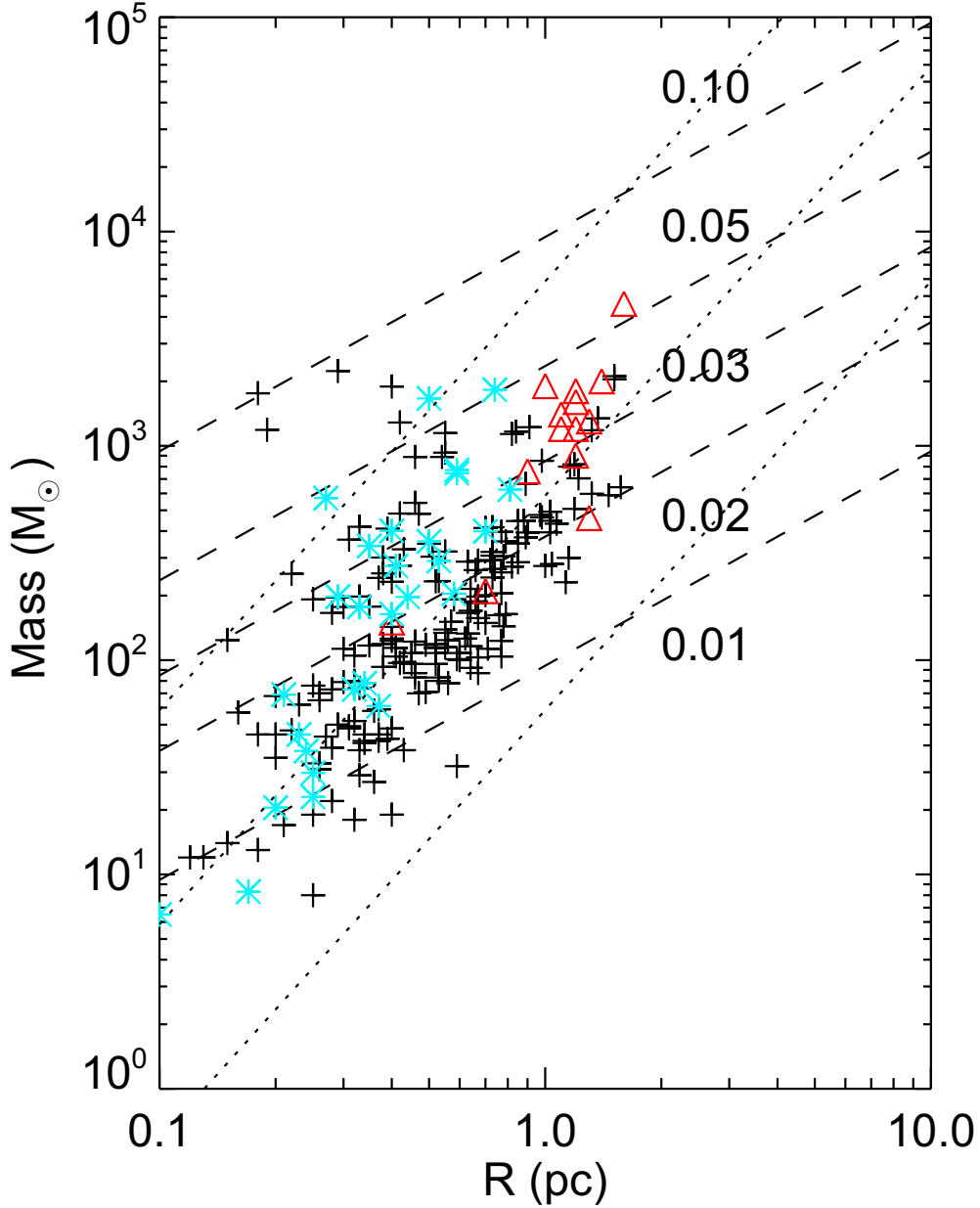}
\caption{Contours of star formation rate per free-fall time predicted by
the outflow-regulated cluster formation model on the mass-radius
diagram. The dashed contours are labeled by values of SFR$_{\rm ff}$.
The dotted  contours indicate the constant column density
at $10^{21}$ cm$^{-2}$ ({\it bottom line}), $10^{22}$ ({\it intermediate
 line}), 
and $10^{23}$ cm$^{-2}$ ({\it upper line}).
The crosses and triangles indicate the cluster-forming dense clumps 
identified by \citet{ridge03} and \citet{higuchi09} in C$^{18}$O
($J=1-0$), respectively.
The asterisks indicate the dense clumps in the infrared dark clouds
identified by \citet{rathborne06}.
For most of the dense clumps, the predicted SFR$_{\rm ff}$ ranges 
from 1 \% to 5 \%, indicative of slow star formation.
}  
\label{fig:SFR}
\end{figure}

Figure \ref{fig:SFR} shows the dependence of SFR$_{\rm ff}$ on the mass and radius, 
obtained from the above equation.
The crosses and diamonds indicate the cluster-forming clumps observed 
in the C$^{18}$O ($J=1-0$) line
by \citet{ridge03} and \citet{higuchi09}, respectively.
Our model suggests that the star formation
rate per free-fall time ranges from 1 \% to 5 \% 
for the observed cluster-forming clumps in the solar neighborhood
when the protostellar outflow feedback maintains the supersonic turbulence 
in the clumps, indicating that it takes about (2 $-$ 10) $t_{\rm ff}$
for the star formation efficiency to reach about (10 $-$ 20) \%.

\clearpage

%%%%%%%%%%%%%%%%%%%%%%%%%%%%%%%%%%%%%%%%%%%%%%%%%%%%%%%%%%%%%%%%%%%%%
%%%%%%%%%%%%%%%%%%%%%%%%%%%%%%%%%%%%%%%%%%%%%%%%%%%%%%%%%%%%%%%%%%%%%
\begin{deluxetable}{llll}
\tabletypesize{\scriptsize}
%\rotate
\tablecolumns{4}
\tablecaption{Model Characteristics}
%\tablewidth{\columnwidth}
\tablewidth{9cm}
\tablehead{\colhead{Model} &\colhead{$\beta$} 
&\colhead{$f$\tablenotemark{a}} 
&\colhead{Remark} 
}
\startdata
N1  & $\infty$ & 0.4 & no B field \\
N2  & $\infty$ & 0.0 & no B field, no feedback \\
%EW1  & 20   & 0.0  & extremely-weak B field  \\
%EW  & 20   & 0.4 &  extremely-weak B field  \\
W1  & 2   & 0.4 &  weak B field  \\
W2  & 2   & 0.0 &  weak B field, no feedback  \\
S1  & 0.2    & 0.4  & moderately-strong B field  \\
S2  & 0.2    & 0.0  & moderately-strong B field, no feedback \\
S3  & 0.2    & 0.1 & moderately-strong B field, weak feedback 
%S4  & 0.2  & 0.4 & on & ambipolar diffusion  
%\hline
\enddata
\tablenotetext{a}{outflow strength}
%\tablenotetext{b}{ambipolar diffusion}
%\tablecomments{}
\label{tab:model}
\end{deluxetable}

\begin{deluxetable}{lllll}
%\tabletypesize{\scriptsize}
%\rotate
\tablecolumns{5}
\tablecaption{Summary of the Physical Properties of the identified 
cores\label{tab:physprop}}
\tablewidth{\columnwidth}
\tablehead{\colhead{Property} & \colhead{Minimum} & \colhead{Maximum}
&\colhead{Mean\tablenotemark{a}}   &\colhead{Median} 
}
\startdata
model N1 &  &  &  & \\ \hline
$R_{\rm c}$ (pc) & 0.013 & 0.041  & 0.020 $\pm$ 0.005  &
0.020 \\
$M_{\rm c}$ (M$_\odot$) & 0.075 & 3.51  & 0.67 $\pm$ 0.67  & 0.40 \\
$\Delta V_{\rm NT}$ (km s$^{-1}$) & 0.13 & 4.53  & 1.06 $\pm$ 0.77  & 0.86 \\ \hline
model W1 &  &  &  &  \\ \hline
$R_{\rm c}$ (pc) & 0.013 & 0.034  & 0.020 $\pm$ 0.006  &
0.019 \\
$M_{\rm c}$ (M$_\odot$) & 0.076 & 3.90  & 0.62 $\pm$ 0.77  & 0.31 \\
$\Delta V_{\rm NT}$ (km s$^{-1}$) & 0.13 & 14.4  & 0.99 $\pm$ 1.66  & 0.47 \\ \hline
model S1 &  &  &  &  \\ \hline
$R_{\rm c}$ (pc) & 0.013 & 0.048  & 0.024 $\pm$ 0.008  &
0.023 \\
$M_{\rm c}$ (M$_\odot$) & 0.073 & 4.39  & 0.86 $\pm$ 1.03  & 0.47 \\
$\Delta V_{\rm NT}$ (km s$^{-1}$) & 0.086 & 2.64  & 0.46 $\pm$ 0.52  & 0.31  
\enddata
\tablenotetext{a}{With standard deviation}
\tablecomments{The gas temperature is assumed to be $T=20$ K,
corresponding to the sound speed of 0.266 km s$^{-1}$.
} 
\end{deluxetable}

\clearpage

\begin{thebibliography}{99}
\bibitem[Andr\'e et al.(2007)]{andre07}
Andr\'e, P., Belloche, A., Motte, F., \& Peretto, N. 2007, \aap,
               472, 519
\bibitem[Arce et al.(2010)]{arce10}
Arce, H. G., Borkin, M. A., Goodman, A. A., Pineda, J. E., \&
Halle, M. W. 2010, ApJ, in press
\bibitem[Ballesteros-Paredes et al.(2009)]{ballesteros09}
Ballesteros-Paredes, J., Gomez, G. C., Pichardo, B., \& 
V\'azquez-Semadeni, E. 2009, \mnras, 393, 1563
\bibitem[Bertoldi \& McKee(1992)]{bertoldi92}
Bertoldi, F. \& McKee, C. F. 1992, \apj, 395, 140
\bibitem[Bontemps et al.(2001)]{bontemps01}
Bontemps, S. 2001, \aap, 372, 173
\bibitem[Butler \& Tan(2009)]{butler09}
Butler, M. J., \& Tan, J. C. 2009, \apj, 696, 484
\bibitem[Carroll et al.(2009)]{carroll09}
Carroll, J. J., Frank, A., Blackman, E. G., Cunningham, A. J., \&
		Quillen, A. C. 2009, \apj, 695, 1376
\bibitem[Crutcher(1999)]{crutcher99}
Crutcher, R. M., 1999, \apj, 520, 706
\bibitem[Davis et al.(2000)]{davis00}
Davis, C. J., Chrysostomou, A., Matthews, H. E., Jenness, T., \& Ray,
		T. P. 2000, \apj, 530, L115
M. 2007, \apj, 61, 262
\bibitem[Dib et al.(2007)]{dib07}
Dib, S., Kim, J., V\'azquez-Semadeni, E., Burkert, A., \& Ahadmehri,
M. 2007, \apj, 61, 262
\bibitem[Di Francesco et al.(2004)]{james04}
Di Francesco, J., Andre, P., \& Myers, P. C. 2004, \apj, 617, 425
\bibitem[Elmegreen(2000)]{elmegreen00}
Elmegreen, B. G. 2000, \apj, 530, 277
\bibitem[Federrath et al.(2011)]{federrath11}
Federrath, C., Sur, S., Schleicher, D. R. G., Banerjee, R., \& Klessen,
		R. S. 2011, \apj in press (axXiv:1102.0266). 
\bibitem[Friesen et al.(2010)]{friesen10}
Friesen, R. K., Di Francesco, J., Shimajiri, Y., \& Takakuwa, S., 2010,
               \apj, 708, 1002
\bibitem[Girart et al.(2009)]{girart09}
Girart, J. M., Beltr\'an, M. T., Zhang, Q., Rao, R., \& Estalella,
		R. 2009, Science, 324, 1408
\bibitem[Hartmann et al.(2001)]{hartmann01}
Hartmann, L. W., Ballesteros-Paredes, J., Bergin, E. A. 2001, \apj, 562, 852
\bibitem[Hennebelle et al.(2003)]{hennebelle03}
Hennebelle, P., Whitworth, A. P., Gladwin, P. P., \& Andr\'e, P. 2003,
               \mnras, 340, 870
\bibitem[Higuchi et al.(2009)]{higuchi09}
Higuchi, A. E., Kurono, Y., Saito, M., \& Kawabe, R. 2009, \apj, 705, 468
\bibitem[Hirota et al.(2008)]{hirota08}
Hirota, T. et al. 2008, \pasj, 60, 37
\bibitem[Houde et al.(2004)]{houde04}
Houde, M. et al. 2004 \apj, 604, 717
\bibitem[Johnstone et al.(2000)]{johnstone00}
Johnstone, D., Wilson, C. D., Moriarty-Schieven, G., et al. 2000, \apj,
545, 327
\bibitem[Krumholz \& McKee(2005)]{krumholz05}
Krumholz, M. R. \& McKee, C. F. 2005, \apj, 630, 250
\bibitem[Klessen et al.(2000)]{klessen00}
Klessen, R., Heitsch, F., \& Mac Low, M.-M. 2000, \apj, 535, 887
\bibitem[Krumholz \& Tan(2007)]{krumholz07}
Krumholz, M. R. \& Tan, J. C., 2007, \apj, 656, 959
\bibitem[Larson(1981)]{larson81}
Larson, R. B. 1981, \mnras, 194, 809
\bibitem[Loinard et al.(2008)]{loinard08}
Loinard, L., Torres, R. M., Mioduszewski, A. J. \& Rodriguez,
         L. F. 2008, \apj, 675, L29
\bibitem[Lombardi et al.(2008)]{lombardi08}
Lombardi, M., Lada, C. J.  \& Alves, J. 2008, \aap, 480, 785
\bibitem[Li \& Nakamura(2006)]{li06}
Li, Z.-Y. \& Nakamura, F. 2006, \apj, 640, L187
\bibitem[Li et al.(2010)]{li10}
Li, Z.-Y., Wang, P., Abel, T., \& Nakamura, F. 2010, \apjl, 720, 26
\bibitem[Loren(1989)]{loren89}
Loren, R. T. 1989, \apj, 338, 902
\bibitem[Mac Low et al.(1998)]{maclow98}
Mac Low, M.-M., Klessen, R. S., Burkert, A., \& Smith, M. D. 
1998, \prl, 80, 2754
\bibitem[Mac Low(1999)]{maclow99}
Mac Low, M.-M. 1999, \apj, 524, 169
\bibitem[Maruta et al.(2010)]{maruta10}
Maruta, H., Nakamura, F., Nishi, R., Ikeda, N., Kitamura, Y.
2010, \apj, 714, 680
\bibitem[Matsumoto et al.(2006)]{matsumoto06}
Matsumoto, T., Nakazato, T., \& Tomisaka, K. 2006, \apj, 637, L105
\bibitem[Matthews et al.(2009)]{matthews09}
Matthews, B. C., McPhee, C. A., Fissel, L. M., \& Curran, R. L. 2009, 
\apjs, 182, 143
\bibitem[Matzner \& McKee(2000)]{matzner00}
Matzner, C. D., \& McKee, C. F. 2000, \apj, 545, 364
\bibitem[Matzner(2007)]{matzner07}
Matzner, C. D. 2007, \apj, 659, 1394
\bibitem[Maury et al.(2009)]{maury09}
Maury A., Andr\'e, P., \& Li, Z.-Y. \apj, 2009, 499, 175
\bibitem[McKee \& Zweibel(1992)]{mckee92}
McKee, C. F., \& Zweibel, E. G. 1992, \apj, 399, 551
\bibitem[Motte et al.(1998)]{motte98}
Motte, F., Andr\'e, P., \& Neri, R. 1998, \aap, 336, 150
\bibitem[Myers(2009)]{myers09}
Myers, P. C. 2009, \apj, 700, 1609
\bibitem[Nakamura \& Li(2007)]{nakamura07}
Nakamura, F. \& Li, Z.-Y. \apj, 2007, 662, 395
\bibitem[Nakamura \& Li(2008)]{nakamura08}
Nakamura, F. \& Li, Z.-Y. \apj, 2008, 687, 354
\bibitem[Nakamura et al.(2011a)]{nakamura11a}
Nakamura, F. et al. 2011a, \apj, 726, 46
\bibitem[Nakamura et al.(2011b)]{nakamura11b}
Nakamura, F. et al. 2011b, \apj, in press (arXiv:1105.4481)
\bibitem[Nakano \& Nakamura(1978)]{nakano78}
Nakano, T., \& Nakamura, T. 1978, \pasj, 30, 681
\bibitem[Ostriker et al.(2001)]{ostriker01}
Ostriker, E. C., Stone, J. M., \& Gammie, C. F., 2001, \apj, 546, 980
\bibitem[Padoan et al.(2001)]{padoan01}
       Padoan, P., Goodman, A., Draine, B. T., Juvela, M., 
Nordlund, A., \& Rognvaldsson, O. E. 2001, \apj, 559, 1005
\bibitem[Padoan \& Nordlund(2002)]{padoan02}
       Padoan, P. \& Nordlund, A. 2002, \apj, 576, 870
\bibitem[Pelkonen et al.(2007)]{pelkonen07}
       Pelkonen, V.-M., Juvela, M., \& Padoan, P. 2007, \aap, 461, 551
\bibitem[Price \& Bate(2008)]{price08}
Price, D. J., \& Bate, M. R. 2008, \mnras, 385, 1820
\bibitem[Price \& Bate(2009)]{price09}
Price, D. J., \& Bate, M. R. 2009, \mnras, 398, 33
\bibitem[Rathborne et al.(2003)]{rathborne06}
Rathborne, J. M., Jackson, J. M., \& Simon, R. 2006, \apj, 641, 389
\bibitem[Ridge et al.(2003)]{ridge03}
Ridge, N. A., Wilson, T. L., Megeath, S. T., Allen, L. E., \& 
Myers, P. C. 2003, \aj, 126, 286
\bibitem[Smith et al.(2008)]{smith08}
Smith, R. J., Clark, P. C., \& Bonnell, I. A. 2008, \mnras, 391, 1091
\bibitem[Stanke et al.(2006)]{stanke06}
Stanke, T., Smith, M. D., Gredel, R., \& Khanzadyan, T. 2006, \aap, 447, 609
\bibitem[Sugitani et al.(2010)]{sugitani10}
Sugitani, K., Nakamura, F., Tamura, M., Watanabe, M., Kandori, R.,
               Nishiyama, S., Kusakabe, N., Hashimoto, J., Nagata, T.,
               \& Sato, S. 2010, \apj, 716, 299
\bibitem[Sugitani et al.(2011)]{sugitani11}
Sugitani, K. et al. 2011, \apj, 734, 63
\bibitem[Testi \& Sergent(1998)]{testi98}
Testi, L. \& Sergent, A. 1998, \apj, 508, L91
\bibitem[Tilley \& Pudritz(2007)]{tilley07}
Tilley, D., \& Pudritz, R. 2007, \mnras, 382, 73
\bibitem[Walsh et al.(2007)]{walsh07}
Walsh, A. J., Myers, P. C., Di Francesco, J., Mohanty, S., Bourke,
               T. L., Gutermuth, R. \& Wilner, D. 2007, \apj, 655, 958
\bibitem[Wang et al.(2010)]{wang10}
Wang, P., Li, Z.-Y., Abel, T., \& Nakamura, F. 2010, \apj, 709, 27
\bibitem[Williams et al.(1994)]{williams94}
Williams, J. P., de Geus, E. J., \& Blitz, L. 1994, \apj, 428, 693
\end{thebibliography}
\end{document}